\documentclass[12pt]{article}
\usepackage{graphicx, epsfig, color}
\usepackage{amssymb,amsmath}

\def\eslt{E_T^{\rm miss}}
\def\delew{\Delta_{\rm EW}}

\def\delhs{\Delta_{\rm HS}}
\def\delbg{\Delta_{\rm BG}}
\def\to{\rightarrow}

\def\bi{\begin{itemize}}
\def\ei{\end{itemize}}

\def\tb{\tilde b}

\def\tst{\tilde t}

\def\tg{\tilde g}

\def\tw{\widetilde W}
\def\tz{\widetilde Z}

\def\be{\begin{equation}}  
\def\ee{\end{equation}}  
\def\bea{\begin{eqnarray}}  
\def\eea{\end{eqnarray}}  
\def\beas{\begin{eqnarray*}}  
\def\eeas{\end{eqnarray*}}

\newcommand\prd[3]{{\it Phys.\ Rev.\ }{\bf D #1} (#2) #3}

\newcommand\prl[3]{{\it Phys.\ Rev.\ Lett.\ }{\bf #1} (#2) #3}
\newcommand\plb[3]{{\it Phys.\ Lett.\ }{\bf B #1} (#2) #3}
\newcommand\jhep[3]{{\it J. High Energy Phys.\ }{\bf #1} (#2) #3}

\newcommand\epjc[3]{{\it Eur.\ Phys.\ J. }{\bf C #1} (#2) #3}

\begin{document}
\begin{titlepage}
\begin{flushright}
UH-511-1245-15
\end{flushright}

\vspace{0.5cm}
\begin{center}
{\Large \bf  Supersymmetry: Aspirations and Prospects
}\\ 
\vspace{1.2cm} \renewcommand{\thefootnote}{\fnsymbol{footnote}}
{\large 
Xerxes Tata$^1$\footnote[1]{Email: tata@phys.hawaii.edu } 
}\\ 
\vspace{0.5cm} \renewcommand{\thefootnote}{\arabic{footnote}}
  
{\it 
$^1$Dept. of Physics and Astronomy,
University of Hawaii, Honolulu, HI 96822, USA \\
}
\end{center}

\vspace{0.5cm}
\begin{abstract}
\noindent 
\vspace*{0.3cm}

 The realization in the early 1980s that weak scale supersymmetry
stabilizes the Higgs sector of the spectacularly successful Standard
Model led several authors to explore whether low energy supersymmetry
could play a role in particle physics. Among these were Richard
Arnowitt, Ali Chamseddine and Pran Nath who constructed a viable {\em
locally} supersymmetric Grand Unified Theory (GUT), laying down the
foundation for supergravity GUT models of particle physics.
Supergravity models continue to be explored as one of the most promising
extensions of the Standard Model. After a quick overview of some of the
issues and aspirations of early researchers working to bring
supersymmetry into the mainstream of particle physics, we re-examine
early arguments that seemed to imply that superpartners would be
revealed in experiments at LEP2 or at the Tevatron.  Our purpose is to
assess whether the absence of any superpartners in searches at LHC8
presents a crisis for supersymmetry. Toward this end, we re-evaluate
fine-tuning arguments that lead to upper bounds on (some) superpartner
masses. We conclude that phenomenologically viable superpartner spectra
that could arise within a high scale model tuned no worse than a few
percent are perfectly possible. While no viable underlying model of
particle physics that leads to such spectra has yet emerged, we show
that the (supergravity-based) Radiatively-driven Natural Supersymmetry
(RNS) framework serves as a surrogate for a phenomenological analysis of
an underlying theory with modest fine-tuning. We outline the
phenomenological implications of this framework, with emphasis on those
LHC and electron-positron collider signatures that might point to the
underlying natural origin of gauge and Higgs boson masses. We conclude
that the supergravity GUT paradigm laid down in 1982 by Arnowitt,
Chamseddine and Nath, and others, remains a vibrant possibility.

\end{abstract}

\end{titlepage}

\section{Introduction}

\subsection{Historical Prelude}
Supersymmetry (SUSY) phenomenology has been an active area of research since
the early 1980s. The direct search for the superpartners has been one of
the central items on the agenda of $e^+e^-$, $ep$ and hadron collider
experiments at the energy frontier for over three decades now. In
addition, there are (and have been) many experiments operating at lower
energies that are also searching for quantum effects of 
supersymmetric particles that would modify the properties of quarks and
leptons; {\it e.g.} rare decays of bottom mesons, or the magnetic moment
of the muon.\footnote{Indeed some ``true believers'' of SUSY
have gone out on a limb and interpreted the fact that the measured
values of gauge couplings measured at LEP appear to unify in the Minimal
Supersymmetric Standard Model (MSSM) but not in the Standard Model as
evidence for the virtual corrections from sparticles. While Grand
Unification is a very pretty idea, we have to keep in mind there is, as
yet, no direct evidence for it.} Finally, searches
for dark matter are often  interpreted
in the context of supersymmetric models. 

That SUSY has become a part of the mainstream of particle physics is the
result of the pioneering efforts of many people, including Richard
Arnowitt and his collaborators.\footnote{See Ref.~\cite{nath} for a
detailed account of this.} Space-time supersymmetry -- to be
distinguished from the conceptually different notion of supersymmetry on
the two-dimensional worldsheet of string theory \cite{wssusy} -- was
discovered as far back as 1971 when Golfand and Likhtman \cite{golfand}
introduced the supersymmetric extension of the Poincar\'e algebra, and
independently by Volkov and Akulov \cite{volkov} who interpreted the
(massless) neutrino as the Goldstone fermion in a model with SUSY
realized non-linearly. The seminal work of Wess and Zumino \cite{wessz}
(who were unaware of the earlier 1971 work just mentioned) presented a
four-dimensional model of a quantum field theory with supersymmetry
realized linearly.  SUSY, however, mostly remained a playground for
quantum field theorists throughout the 1970s. Research in SUSY included,
among other things, efforts to study {\em locally supersymmetric} field
theories as described below.  With the exception of several pioneering
papers by Pierre Fayet \cite{fayet}, no one seemed to connect SUSY with
particle physics. This situation changed dramatically when it was
recognized \cite{hier} that the remarkable ultra-violet properties of
supersymmetric theories would protect the scalar sector of the Standard
Model (SM) from enormous quantum corrections that are generically
present \cite{susswein} when the SM is embedded in a framework with a
hierarchically different mass scale,  as for example, in a Grand Unified
Theory (GUT).
Without supersymmetry, the corrections to the Higgs boson mass squared
parameter which are typically a loop factor times $M_{\rm GUT}^2 \sim
10^{28}$~GeV$^2$ and need to be cancelled to more than twenty significant
figures against the corresponding Lagrangian parameter in order for the
physical Higgs boson mass to be below its unitarity limit of $\sim
600-800$~GeV.  While such a precise cancellation is technically
always possible, the need for it is generally regarded as a flaw in the theory,
and often referred to as the {\em big hierarchy problem}.  The key
observation was that SUSY continues to protect the SM scalar sector even
if it is spontaneously broken (in this case, all SUSY breaking operators
are soft \cite{girardello}), provided that superpartners of SM particles
(at least those with significant couplings to the Higgs sector) are
lighter than a few TeV.

This led many authors to construct globally supersymmetric models of
particle physics with SUSY broken spontaneously \cite{oraif,FI} below
the TeV scale. A feature of a class of these models (at least those
without $U(1)$ factors in the gauge group, as in all models with Grand
Unification) was a sum rule that implied that the sum of squared masses
over all chiral fermionic degrees of freedom had to be equal to the
corresponding sum over all the corresponding bosonic degrees of freedom
\cite{sumrule}. Moreover, the sum rule applied separately in each
electric charge and colour sector of the theory. Assuming that there
are no unknown charge 1/3 quarks, it implied that one of the charge 1/3
superpartners had to be lighter than $m_b$, which was experimentally
excluded. The obvious way out -- introduce new heavy charge 1/3 quarks
-- led to unduly complicated models, as new particles had to be included
in various sectors of the theory.

A phenomenologically viable alternative that side-steps this problem is
to start with a globally supersymmetric $SU(3)\times SU(2)\times U(1)_Y$
Yang-Mills gauge theory with the desired three generations of quarks and
leptons, the $SU(2)$ doublet Higgs fields to spontaneously break the
electroweak symmetry to electromagnetism and, of course, the gauge bosons,
together with the superpartners of the SM particles. SM Yukawa
interactions between the Higgs fields and matter fermions (along with
interactions of the superpartners that are necessary to preserve SUSY)
that are needed to give fermion masses are incorporated via a
superpotential. Since the superpotential has to be a holomorphic
function of the (super)fields (which in plain English means it cannot
include both the field and its Hermitian conjugate), we necessary need two
independent Higgs doublet fields in order to give mass to the up- and
down-type fermions.  Thus, unlike the SM which has just a single
physical Higgs scalar, the simplest supersymmetric model includes five
spin-zero particles in the electroweak breaking sector. We
note that it is possible to include renormalizable gauge-invariant
interactions that violate baryon or lepton number conservation in the
superpotential. Since these can be potentially dangerous, it is
traditional to forbid these by imposing $R$-parity conservation.  As a
final step, supersymmetry breaking is incorporated by including all soft
supersymmetry breaking (SSB) \cite{girardello} terms consistent with
underlying Yang-Mills gauge symmetry and the {\em assumed} $R$-parity
conservation. The resulting theory \cite{mssm} with the minimal particle
content and the fewest number of new interactions necessary for a
phenomenologically viable model of particle physics is called the
Minimal Supersymmetric Standard Model (MSSM).

In parallel with the work that led to particle physics models with an
underlying (broken) global SUSY, starting with the work of Volkov and
Soroka \cite{vs}, field theorists considered the possibility that SUSY,
like Yang-Mills gauge symmetry, is a {\em local} symmetry. The
pioneering efforts of many authors \cite{sugrarev,hist} (including
Arnowitt, Nath and Zumino \cite{ansugra}) culminated in the work of
Cremmer {\it et al.} \cite{cremmer} who, building upon their earlier
work on supergravity couplings of a single chiral supermultiplet
\cite{cremmerearly}, successfully coupled an {\em arbitrary number} of
matter and gauge fields in a locally symmetric manner.\footnote{Many
years back, I had learnt from R.  Arnowitt that R. Arnowitt,
A. Chamseddine and P. Nath had independently obtained supergravity
couplings with arbitrary number of matter multiplets that they had
needed for development of locally supersymmetric particle physics models
\cite{acnbook}. Indeed, they had made crucial use of these results in
their classic 1982 paper \cite{acnsugra}. See also Ref.~\cite{nath}.} The
resulting supersymmetric theory \cite{cremmer,acnbook} included not only
gauge interactions of matter particles, but also interactions of matter
and gauge fields and their superpartners with gravity! It also included
the gravitino, the spin $3/2$ superpartner of the graviton, which after
the spontaneous breaking of the local SUSY, acquired a mass in much the
same way that the gauge bosons acquire mass when local Yang-Mills gauge
symmetry is spontaneously broken.

The introduction of local SUSY had another remarkable consequence. It
led to a modification \cite{cremmer} of the mass sum rule of global SUSY
mentioned earlier by the addition of a positive term proportional to the
squared gravitino mass on the fermion side.\footnote{We refer the
interested reader to Eq. (10.66) of Ref.~\cite{wss}.}  This is exactly
what was needed since now, the superpartners of the matter fermions
could consistently have masses comparable to the gravitino mass which
could be chosen to be at the TeV scale allowing for the construction of
phenomenogically viable supergravity models of particle physics!

\subsection{Supergravity Grand Unification} 

In truly pioneering work which set a new direction for supersymmetric
models of particle physics, R. Arnowitt, A. Chamseddine and P. Nath
(hereafter referred to as ACN) \cite{acnsugra} as well as a number of
other groups \cite{others} worked to create realistic models of particle
physics based on local SUSY \cite{nilles}. The common theme underlying
these models is that a Planck mass field develops a SUSY breaking term
$\langle {\cal F}\rangle$ via its superpotential interactions, and that
the superpartners of SM particles feel the effect of SUSY breaking {\em
only} via their interaction with this field. The novel feature was that
direct (superpotential) couplings of SM particles and their
superpartners with the fields involved with the breaking of SUSY were
forbidden. However, because gravity couples to energy and momentum,
gravitational interactions between these are always present in locally
supersymmetric models. These interactions carry the message of SUSY
breaking to the Standard Model sector, and lead to masses for SM
super-partners. Dimensional analysis requires that these masses must be
$m_{\rm SUSY} \sim \frac{\langle \cal{F} \rangle}{M_P}$, since $m_{\rm
SUSY}$ has to vanish if either $\langle {\cal{F}}\rangle \to 0$ (no SUSY
breaking) or $M_P \to \infty$ (no gravitational interactions), to be
compared with the gravitino mass $m_{3/2} = \frac{\langle
\cal{F}\rangle}{\sqrt{3}M_P}$. ACN developed an $SU(5)$ supergravity GUT
model whose low energy particle content was that of the MSSM augmented
by a singlet Higgs field. Remarkably, they found that the gravitational
interactions that are required by the underlying local SUSY trigger the
spontaneous breakdown of breaking of the $SU(2)\times U(1)_Y \to
U(1)_{\rm em}$ when SUSY is broken, and generate masses for the
electroweak gauge boson masses at the {\em tree level}.

The ACN paper inspired the development of what became the well-known
minimal supergravity (mSugra) model. Points worth noting are: (1)~ACN
assumed a minimal K\"ahler potential in their analysis which led to the
hall-mark universality of {\em all} scalar mass parameters renormalized
at an energy scale $Q\sim M_{\rm GUT}-M_P$.\footnote{The ACN model was
recognized to be an effective theory obtained by integrating out Planck
scale fields in the SUSY-breaking sector.}  It was pointed out
that high scale non-universality can readily be incorporated by allowing
non-universal K\"ahler potential \cite{soniweldon}. Scalar mass
non-universality plays 
an essential role in the model discussed in
Sec.~\ref{sec:spectra}. (2)~ACN included a SM singlet in their analysis in
order to break electroweak symmetry at the tree-level. Several authors
\cite{radewsb} had, however, noted that even if the scalar mass high
scale parameters are universal, top quark Yukawa interactions could
drive the corresponding squared Higgs mass parameter to a negative value
at the weak scale relevant for electroweak phenomenology, {\em provided
that} $m_t \ge 100$~GeV. In other words, the singlet is not required for
electroweak symmetry breaking. (3)~ACN's choice of the minimal form for
the K\"ahler potential caused the bilinear and trilinear soft SUSY
breaking scalar coupling parameters to be simply related which led to a
fixed value for $\tan\beta$. It was, however, recognized very early that
the Higgs sector parameters would likely be modified to accommodate the
phenomenologically required scales for $\mu$ and $b_\mu$, and
$\tan\beta$ was treated as an independent parameter, essentially from
the onset. The mSUGRA model is thus characterized by the well-known
parameter set,
\be
m_0,m_{1/2},A_0,\tan\beta, sign(\mu).
\label{eq:msugra}
\ee
In writing (\ref{eq:msugra}), we have assumed that the gauginos all
arise from a unified gaugino mass parameter because of an underlying GUT
and that the order parameter $\cal{F}$ for SUSY breaking respects the
GUT symmetry, and also that the magnitude of $\mu$ (but not its
sign\footnote{We are tacitly assuming that $\mu$ is real, or more
precisely, that there are no relative phases between the various
dimensionful parameters that could lead to $CP$ violation in the SUSY
sector.}) is fixed to yield the observed value of $M_Z^2$. The various
superpartner masses are then obtained by evolving these soft parameters
from their universal values at the high scale down to the weak scale
relevant for phenomenology in much the same way that the
phenomenologically relevant QCD and electroweak gauge couplings are
obtained from a single gauge coupling in a GUT.

\subsection{Aspirations of Supersymmetry} \label{subsec:aspir}

As we mentioned above, the realization that SUSY could stabilize the
Higgs sector in the presence of radiative corrections led to an
explosion of activity to devise strategies by which SUSY would reveal
itself in various experiments.  Although this may seem odd today, in the
1980s it was  standard practice to motivate the examination of
supersymmetric extensions of the SM, and typically a paper on SUSY
phenomenology would begin by noting that:
\begin{enumerate}
\item Supersymmetry is the largest possible space-time symmetry of the
$S$-matrix;
\item Supersymmetry provides a synthesis of bosons and fermions never
  previously attained; 
\item Local supersymmetry provides a connection to gravity;
\item Assuming $R$-parity conservation (which was motivated by
  considerations of proton stability) supersymmetry naturally provides a
  candidate for particle dark matter;
\item Weak scale supersymmetry solves the {\it big hierarchy problem} in
that (in a softly broken supersymmetric theory) low scale physics does not
exhibit quadratic sensitivity to physics at high scales; for
instance, when the MSSM is embedded into a framework with a
hierarchically separated  mass scale
such as a SUSY GUT.
\end{enumerate}
While each of these items provides strong motivation for studying
supersymmetric theories, it is worth stressing that it is just the last
item that provides motivation for superpartners at the TeV scale
relevant to supercolliders such as the LHC. It was indeed exciting that
the measured values of the gauge couplings at LEP (at the end of the
1980s) were compatible with Grand Unification in the MSSM but not in the
SM provided that superpartner masses were in 0.1-10~TeV range \cite{gaugeunif}!

\section{The Mass Scale of Superpartners: An Introspection}

Since the  stability of the electroweak symmetry breaking sector
plays a central
role in our considerations about what might lie beyond the SM, it seems
worthwhile to re-assess the arguments that led us to infer the 
existence of new physics close to the weak scale.
In a generic quantum field theory, the squared  mass of a scalar boson
(such as the Higgs boson of the SM) is given in terms of the
corresponding Lagrangian parameter, $m_{\phi 0}^2$, by,  
\be
m_{\phi}^2 =  m_{\phi 0}^2 + C_1 \frac{g^2}{16\pi^2}\Lambda^2 + 
C_2 \frac{g^2}{16\pi^2}m_{\rm low}^2 \log\left(\frac{\Lambda^2}{m_{\rm low}^2}\right) 
+C_3 \frac{g^2}{16 \pi^2}m_{\rm low}^2\;.
\label{eq:generic}
\ee 
Here, $\Lambda$ denotes the scale up to which the effective theory
which is being used to evaluate the scalar mass is valid, $m_{\rm low}$
is the highest mass scale of the low energy theory, $g$ denotes a
typical coupling constant and the $C_i$ are dimensionless numbers
(including spin
and multiplicity factors) typically ${\cal O}(1)$.  The $C_3$ term could
also include ``small logarithms'' $\log(m_{\rm low}^2/m_\phi^2)$ that we
have not exhibited. For instance, if the low energy theory is the SM
viewed as embedded in a GUT, $m_{\rm low}$ will
be about $M_Z$, $m_h$, $m_t$ or the SM vacuum expectation value $v$, while
$\Lambda$ will be $M_{\rm GUT}$, since the SM becomes invalid for energy
scales higher than $M_{\rm GUT}$ because GUT scale degrees of freedom
({\it e.g.} the GUT and coloured Higgs bosons with masses around $M_{\rm
GUT}$) are not included in the SM. In the case of the MSSM embedded in a
SUSY GUT, $m_{\rm low} \sim m_{\rm SUSY}$ with $\Lambda \sim M_{\rm
GUT}$. The $C_1$ term which is quadratic in $\Lambda$ is what leads to
the {\it big hierarchy problem} that destabilizes the SM if it is
embedded into a theory with very heavy particles such as a
GUT.\footnote{We emphasize that $\Lambda$ in Eq.~(\ref{eq:generic}) is
{\em not a regulator} associated with divergences that occur in quantum
field theory but is a physical scale associated with new particles that
have significant couplings to the scalar sector. In other words, the
terms in Eq.~(\ref{eq:generic}) are the finite terms (after
renormalization) that would result from a calculation using the high scale
theory. For this same reason, although it is tempting, we refrain from
choosing $\Lambda \sim M_{\rm Planck}$; we do not know quantum
gravitational dynamics, and in particular, cannot say that there are
associated new particles with large couplings to the Higgs sector of the
SM; see also Ref.~\cite{drees}.} Because the $C_1$ term is never present in
softly broken SUSY, the {\it big hierarchy problem} is automatically
solved, as long as $m_{\rm SUSY}\ll M_{\rm GUT}$. 

Applying Eq.~(\ref{eq:generic}) to the MSSM embedded in a GUT, we see
that the leading correction to the Higgs boson mass is given by the
$C_2$ term; {\it i.e.} $$\delta m_h^2 \sim C_2
\frac{g^2}{16\pi^2}m_{\rm SUSY}^2 \log\left(\frac{M_{\rm
    GUT}^2}{m_{\rm low}^2}\right).$$ In the early days, many authors
argued that in order not to have unnatural cancellations, it is
reasonable to set $\delta m_h^2 \lesssim m_h^2$. Indeed, $\Delta_{\rm
  log} = {\delta m_h^2 \over m_h^2}$ was proposed as a simple measure
of the degree of fine-tuning, and continues to be used
by several authors \cite{knpap}. Since the logarithm $\sim 30-40$ if
$m_{\rm SUSY}$ is near the TeV scale, this lead to the conclusion
$m_{\rm SUSY}^2 \lesssim m_h^2$, strongly suggesting that
superpartners would be discovered either at LEP2 or at the
Tevatron. We all know that things did not turn out this way, and it
behooves us to re-examine these arguments more closely in order to
assess whether we should remain optimistic about weak scale SUSY.
With this in mind, we note that:
\bi
\item Perhaps, $\delta m_h^2 < m_h^2$ is too stringent a requirement; we
  know many examples of accidental cancellations in nature of one or two
  orders of magnitude.\footnote{The well-known $\pi^2-9$ factor in the
  decay rate of orthopositronium is a cancellation of one order of
  magnitude. Even more familiar, and perhaps more mysterious, is the fact
  that the angular sizes of the sun and the moon are the same to within
  3\%!}

\item It has long been emphasized that the arguments we made really
  apply only to those superpartners with large couplings to the Higgs
  sector, and so do not apply to first (or even second generation)
  squarks and gluinos whose masses are most stringently probed at the
  LHC. These superpartners couple to the Higgs sector only at two-loop
  so that their masses could easily be $\sim 5-10$~TeV or more
  because there would be an additional $16\pi^2$ in the $C_2$ term of
  Eq.~(\ref{eq:generic}).\footnote{We mention that the
  $D$-term coupling contributions  cancel.}

\item Most importantly, the argument that led us to suggest $\Delta_{\rm
  log}$ as a measure of the degree of cancellation in
  Eq.~(\ref{eq:generic}) assumes that {\em contributions from various
  superpartners are  independent}. It seems almost certain that we
  will find that various superpartner masses are correlated once we
  understand the mechanism of supersymmetry breaking so that {\em
  automatic cancellations} between contributions from various
  superpartners could well occur when we evaluate the fine-tuning in any
  high scale theory. {\it Ignoring these correlations, will overestimate
  the ultra-violet sensitivity of any model.}  
\ei  
These correlations
  are most simply incorporated into the most commonly used fine-tuning
  measure introduced by Ellis, Enqvist, Nanopoulos and Zwirner
  \cite{eenz} and subsequently explored by Barbieri and Guidice
  \cite{bg}: 
\be 
\Delta_{\rm BG}
\equiv max_i\left|\frac{p_i}{M_Z^2}\frac{\partial M_Z^2}{\partial
  p_i}\right|\;.
\label{eq:DBG}
\ee
Here, the $p_i$'s are the {\em independent} underlying parameters of the
theory.  It does not matter that $M_Z^2$ rather than $m_h^2$ is used to
define the sensitivity measure since essentially both the quantities are
proportional to the square of the Higgs field $vev$.\footnote{Indeed,
the quantity $\delhs$ introduced in 
Ref.~\cite{baersugra} to measure
the sensitivity of $M_Z^2$ in a high scale theory also ignores correlations
between various contributions. In this sense, $\delhs$ is analogous to
$\Delta_{\rm log}$ introduced earlier.} The key difference is that
$\delbg$ here measures the sensitivity with respect to the {\em
independent} parameters of any model and so takes into account the
correlations that we mentioned. Since $\delbg$ ``knows about''
correlations that are ignored in $\Delta_{\rm log}$, we expect (aside
from technical caveats that we will not go into here) $\Delta_{\rm log}
\ge \delbg$, which is why we said $\Delta_{\rm log}$ would over-estimate
the degree of fine-tuning.  

We will see below that it is possible to find spectra where the
fine-tuning {\em may be} as small as a part in ten, but where first
and second generation squarks and gluinos are in the multi-TeV range
and third generation sfermions masses at 1-4~TeV. However, before doing so
we introduce the notion of electroweak fine-tuning and
discuss its utility and limitations.

\subsection{Electroweak Fine-tuning} \label{subsec:ewft}

The value of $M_Z^2$ obtained from the minimization of the
one-loop-corrected Higgs boson potential of the MSSM
\be 
\frac{M_Z^2}{2} = \frac{(m_{H_d}^2+\Sigma_d^d) - (m_{H_u}^2+\Sigma_u^u) \tan^2\beta}{\tan^2\beta -1} -\mu^2,
\label{eq:mZsSig}
\ee 
is the starting point for many discussions of fine-tuning.
Eq.~(\ref{eq:mZsSig}) is obtained using the weak scale
MSSM Higgs potential, with all parameters  
evaluated at the scale $Q=M_{SUSY}$.  The $\Sigma$s in
Eq.~(\ref{eq:mZsSig}), which
arise from one loop corrections to the Higgs potential, are the analogue
of the $C_3$ term in (\ref{eq:generic}). Explicit forms
for the $\Sigma_u^u$ and $\Sigma_d^d$ are given in the Appendix of
Ref.~\cite{rns}.  

Requiring that the observed value of $M_Z^2$ is obtained without large
cancellations means that none of the various terms on the
right-hand-side of Eq.~(\ref{eq:mZsSig}) has a magnitude much larger
than $M_Z^2$.  This then suggests that the electroweak fine-tuning of
$M_Z^2$ can be quantified by $\delew^{-1}$, where
\cite{ltr,baersugra,rns} \be \Delta_{\rm EW} \equiv max_i
\left|C_i\right|/(M_Z^2/2)\;.
\label{eq:delew}
\ee 
Here, $C_{H_d}=m_{H_d}^2/(\tan^2\beta -1)$,
$C_{H_u}=-m_{H_u}^2\tan^2\beta /(\tan^2\beta -1)$ and $C_\mu =-\mu^2$.
Also, $C_{\Sigma_u^u(k)} =-\Sigma_u^u(k)\tan^2\beta /(\tan^2\beta -1)$
and $C_{\Sigma_d^d(k)}=\Sigma_d^d(k)/(\tan^2\beta -1)$, where $k$ labels
the various loop contributions included in Eq.~(\ref{eq:mZsSig}).  We
immediately see that any upper bound on $\delew$ that we impose from
naturalness considerations necessarily implies a corresponding limit on
$\mu^2$, a connection first noted almost two decades ago \cite{CCN}. We
conclude that higgsino masses are necessarily bounded from above in any
theory with small values of $\delew$. There are, however, potental
loopholes in the analysis that led us to infer that higgsinos must be
light that we make explicit.
\bi
\item Our reasoning assumes that the superpotential parameter $\mu$ is
  independent of the SSB parameters. If $\mu$ were correlated to the
  SSB parameters (in particular with $m_{H_u}^2$), there would be
  automatic cancellations that would clearly preclude us from
  concluding that higgsinos are light. The Giudice-Masiero \cite{gm}
  mechanism  for the origin of $\mu$
  notwithstanding, we take the view that the superpotential and SSB
  breaking sectors likely have different physical origin, and so are
  unrelated.

\item We assume that there is no soft SUSY breaking contribution to the
  higgsino mass ({\it i.e.} the $\mu^2$ that enters in
  Eq.~(\ref{eq:mZsSig}) via the scalar Higgs potential is indeed the
  higgsino mass parameter). While it is logically possible to include a SSB
  higgsino mass parameter as long as there are no SM singlets with
  significant couplings to the higgsinos, in all high scale models with
  minimal low energy particle content that we are aware of, higgsino
  masses have a supersymmetric origin. In this connection, we mention
  that Nelson and Roy \cite{nr} and Martin \cite{martin} have
  constructed models with additional adjoint chiral superfields at the
  weak scale in which the parameters in the Higgs boson sector are
  logically independent of higgsino masses.

\item It has been pointed out \cite{ckl} that if the Higgs particle is a
  pseudo-Goldstone boson in a theory with an (almost) exact global
  symmetry, it is possible that the Higgs boson remains light even if
  the higgsinos are heavy because cancellations that lead to a low Higgs
  mass (and concomitantly low $M_Z^2$) are a result of a symmetry. We
  note that the model includes several additional fields to have
  complete multiplets of the global symmetry, which is simply put in by hand. 

\ei
Despite these caveats we find it compelling that in models with a
minimal (low energy) particle content the higgsino mass enters 
Eq.~(\ref{eq:mZsSig}) directly, so that a low value of $\delew$ implies the
existence of higgsinos close in mass to $M_Z$. Since we see no strong
motivation for the introduction of several extra fields at the weak
scale, we will continue to regard the existence of light higgsinos as a
necessary condition for natural SUSY in the rest of this paper.

Before proceeding further, we remark that $\delew$ as defined here
entails only weak scale parameters (see also Ref.~\cite{perel}) and so
has no information about the $\log\Lambda$ terms that cause weak scale
physics to exhibit logarithmic sensitivity to high scale physics.  For
this reason {\em we do not view $\delew$ as a fine-tuning measure in the
underlying high scale theory}, as already noted in
Ref.~\cite{rns}. Indeed precisely because $\delew$ does not contain
information about the large logs, we expect that $$\delew \leq \delbg.$$
We instead  regard $\delew^{-1}$ as the minimum fine-tuning in any theory with
a given superpartner spectrum.

\subsection{The Utility of $\delew$} \label{subsec:util}
 Although it is not a fine-tuning measure of a high scale theory,
 $\delew$ is nonetheless useful for many reasons.  \bi

\item Since it depends only on weak scale parameters, $\delew$ is
  essentially determined by the SUSY spectrum, and so is
  ``measureable'', at least in principle. 

\item $\delew$ gives a bound on the fine-tuning in any theory with a
  given SUSY spectrum. Modulo the caveats discussed above, if $\delew$
  turns out to be large, the underlying theory that yields this
  spectrum will be fine-tuned. While small $\delew$ {\em does not
    imply} the absence of fine-tuning, it leaves open
  the possibility of finding an underlying theory with the same
  superpartner spectrum where
SSB parameters are 
  correlated so that the large logarithms
  automatically (nearly) cancel.\footnote{The possibility that
    correlations among underlying parameter reduces the fine-tuning
    has been noted by other authors \cite{reduce}.}  In this
  underlying theory, $\delbg$ will be numerically close to $\delew$,
  once that the correlations among the SSB parameters are incorporated
  in the evaluation of $\delbg$.\footnote{See Ref.~\cite{am}, Sec. 3
    for a detailed illustration of how the cancellations might occur.}
  We emphasize that evaluation of $\delbg$ is essential to declare
  that a high scale theory is free of fine-tuning, and further, that
  evaulation of both $\delbg$ and $\delew$ are necessary to see if the
  underlying correlations yield the minimum fine-tuning needed for a
  given spectrum.\footnote{There is an obvious technical caveat here:
    if a theory is really determined by a single mass scale and {\it
      has no dimensionless free parameters}, any reasonable
    fine-tuning measure should be unity. That this is not the case for
    $\delew$ is because potential correlations between weak scale
    parameters are ignored in its definition. This caveat is
    irrelevant for practical purposes in most cases (where $M_Z^2$
    depends on at least two independent parameters) and the underlying
    theory is defined at a very high scale.}

\item Many aspects of SUSY phenomenology are determined by the
  superpartner spectrum. An investigation of the phenomenology of models
  with low $\delew$ is in effect an  investigation of the phenomenology 
  of (potentially) natural underlying theories. We should, however, be cautious
  about drawing phenomenological conclusions that are sensitive to
  assumed correlations (over and above those dictated by naturalness) in
  the spectrum.

\item As we saw, low $\delew$ imples $\mu^2$ has to be close to $M_Z^2$,
  but squarks (including $t$-squarks) and gluinos may be relatively
  heavy as we will see shortly.

\ei 
In light of this, and despite the caveats that we mentioned above, we
regard light higgsinos as the most robust feature of natural SUSY models
(at least those with near-minimal low energy particle content), and
focus on the phenomenology of models with small $|\mu|$ and
concomitantly light higgsinos.

\section{Generating Spectra with Low $\delew$}
\label{sec:spectra}

As we have already discussed, the the magnitude of $\mu$ is fixed
using Eq.~(\ref{eq:mZsSig}) which
is well approximated by,
$$
\frac{1}{2}M_Z^2 \simeq -(m_{H_u}^2+\Sigma_u^u) -\mu^2,
$$ for moderate to large $\tan\beta$. Thus, aside from radiative
corrections, a weak scale value of
 $-m_{H_u}^2$ close to $M_Z^2$ guarantees a correspondingly small value
 of $\mu^2$.  Within the mSUGRA model $m_{H_u}^2$ evolves to a negative
 value at the weak scale (this is the celebrated mechanism of radiative
 electroweak symmetry breaking \cite{radewsb}), and its magnitude is
 comparable to that of other weak scale SSB parameters.  The resulting
value of  $\mu^2$ is
 typically much larger than $M_Z^2$ as long as the radiative corrections
 contained in $\Sigma_u^u$ are of modest size, and $\delew$ is
 typically large in the mSUGRA model \cite{baersugra}.

\subsection{Radiatively-driven Natural Supersymmetry (RNS)}

A small weak scale value of $m_{H_u}^2$ can always be guaranteed if we
relax the assumption of high scalar mass parameter universality that is
the hallmark of mSUGRA, and treat the Higgs field mass parameters as
independent of corresponding matter scalar masses. The Non-Universal
Higgs Mass model, which has two additional GUT scale parameters
$m_{H_u}^2$ and $m_{H_d}^2$ (NUHM2 model) over and above the the mSUGRA
parameter set (\ref{eq:msugra}), provides an appropriate setting
\cite{nuhm2}.  As discussed in detail in Ref.~\cite{rns}, the added
parameter freedom in the NUHM2 model allows us to find
phenomenologically viable solutions with $\delew$ as small as 10,
corresponding to electroweak fine-tuning of no worse than 10\%.  We
stress that from the perspective of the NUHM2 framework this
necessitates a fine-tuning in that $\delew$ is very sensitive to the
value of $m_{H_u}^2({\rm GUT})$: see Table 1 of Ref.~\cite{rns}.  This is
reflected in the large value of $\delbg$ for the NUHM2 model points in
this table, even though the corresponding $\delew^{-1}$ is just a few
percent.\footnote{We refer the reader to the first row of Table ~1 of
Ref.~\cite{am}. The subsequent rows of this table show how correlations
among the parameters reduce the value of $\delbg$ until we eventually
obtain $\delbg \simeq \delew$.} For this reason, we regard the NUHM2
model to be fine-tuned.  However, as discussed at length in Sec. 3 of
Ref.~\cite{am}, it still leaves open the possibility of discovering an
underlying theory with essentially the same mass spectrum but where the
SSB parameters $m_{H_u}^2$, $A_0$ and $m_{1/2}$ are correlated with
$m_0$ for which $\delbg^{-1} \simeq \delew^{-1}$ is just a few
percent. Such a theory, if it exists, would not be fine-tuned and would
have essentially the same phenomenology as the NUHM2 model.

To find these low $\delew$ solutions, we performed scans of the NUHM2
parameter space \cite{rns,rnslhc,meas},
requiring that:
\bi

\item electroweak symmetry is radiatively broken, 

\item LEP2 and LHC bounds on superpartner masses are respected, and 

\item the value of $m_h$ is consistent with the value of the Higgs boson
mass measured at the LHC \cite{lhchiggs} within a theoretical error
that we take to be $\sim 3$~GeV.
\ei
The low $\delew$ solutions clearly have low values of $|\mu|$, and
generally have $A_0 \sim -(1-2)m_0$; this range of $A_0$ leads to a
cancellation of the $\tst_1$ contribution to $\Sigma_u^u$ (the
$\tst_2$ contribution is suppressed if $m_{\tst_2} \sim
(2.5-3)m_{\tst_1}$), and at the same time results in large
intra-generational top squark mixing that raises the Higgs mass to
$\sim 125$~GeV. Since the required small value of $|\mu|$ is obtained
by $m_{H_u}^2$ being driven from its GUT scale choice to close to
$-M_Z^2$ at the weak scale, this scenario has been referred to as {\em
  Radiatively-driven Natural Supersymmetry} (RNS).  We view the RNS framework
as a
surrogate for the underlying natural model of supersymmetry, and
urge its use for phenomenological analysis of natural SUSY models.

Requiring (somewhat arbitrarily) that $\delew \leq 30$, the RNS spectrum
is characterized by: 
\bi
\item The presence of four higgsino-like states $\tz_1,\tz_2$ and
  $\tw_1^\pm$ with masses in the  100-300~GeV range, with a mass splitting
$\sim 10-30$~GeV between $\tz_2$ and the lightest supersymmetric
  particle  (LSP).

\item $m_{\tg} \sim 1.5-5$~TeV, with $\tz_{3,4}$ and $\tw_2^\pm$ masses
  fixed by (the assumed) gaugino mass unification condition.

\item $m_{\tst_1}=1-2$~TeV, $m_{\tst_2}, m_{\tb_{1,2}} \sim 2-4$~TeV;
  this is in contrast to other studies that suggest that naturalness
  requires that top squarks should all be in the few hundred GeV range, and
  so likely be accessible at the LHC \cite{knpap}. The difference arises
  because we allow for the possibility that SSB parameters may be
  correlated.

\item First and second generation sfermions in the tens of TeV
  range. This is not required to get low $\delew$, but compatible
  \cite{intra} with it. Sfermion masses in this range ameliorate the
  SUSY flavour and CP problems \cite{flavdec}, and also raise the proton
  lifetime \cite{proton}.

\ei

We stress that attaining a small value of $\delew$ (in a manner
consistent with phenomenological constraints) is not trivial in high
scale models. Within the mSUGRA framework $\delew \ge 200$
\cite{baersugra} because we need to be in the so-called
hyperbolic branch/focus point region (HB/FP) \cite{CCN,fp} in order to
generate the required small, negative weak scale value of
$m_{H_u}^2$. This, however, requires multi-TeV values of $m_0$ and
results in relatively large contributions to $\Sigma_u^u$ from loops
involving top-squarks. Indeed Baer {\it et al.} \cite{siege} have
emphasized that of the many high scale SUSY models that have been
considered in the literature, only the NUHM2 model allows for spectra
with $\delew \le 30$ that we have adopted as a necessary criterion for
naturalness.

\subsubsection{Fine-tuning }

Before closing this section, we stress that our interpretation of
$\delew$ differs sharply from that in Ref.~\cite{baervol} where it is
argued that $\delew$, correctly used, is the appropriate measure of
fine-tuning.  These authors use the RNS framework to illustrate their
argument. They note that in a gravity-mediated SUSY breaking framework,
the various soft-SUSY-breaking parameters (but not $\mu$) are all
proportional to the gravitino mass with proportionality constants
$\sqrt{a_i}$, leading them to rewrite the expression for $M_Z^2$ in
terms of {\em high scale} parameters (used for evaluating $\delbg$) as,
\be
M_Z^2 = a m_{3/2}^2 -2.18\mu^2_{\rm GUT}.
\label{eq:fake}
\ee
Here $a$ is the coefficient of $m_{3/2}^2$ that results after {\em
combining} the various contributions from the SSB terms to $M_Z^2$, and
2.18 arises from the (small) renormalization of the
$\mu$-parameter. They then note that for low $\delbg$, $\mu_{\rm GUT}^2$
and $am_{3/2}^2$ must {\em both} be comparable to $M_Z^2$, and argue
that in this framework $\delbg$ automatically approaches $\delew$ {\em
because all SSB parameters are written in terms of the single parameter
$m_{3/2}^2$.} We would agree with this reasoning if we had a theory that
fixed the various coefficents $a_i$ so that there was an automatic
cancellation that resulted in $a m_{3/2}^2 \sim M_Z^2$. In the
gravity-mediated scenario envisioned in Ref.~\cite{baervol}, however, the
coefficients $a_i$ are fixed in terms of the parameters of the hidden
sector superpotential, K\"ahler potential and gauge kinetic
functions. In the absence of an underlying theory of hidden sector
physics, we regard Eqs.~(31-35) of Ref.~\cite{baervol} simply as a
re-parametrization of the MSSM SSB parameters, which cannot alter any
conclusions as to whether or not the theory is fine-tuned! Stated
differently, we would agree with the authors of Ref.~\cite{baervol} that
the fine-tuning measure reduces to $\delew$ in the RNS framework if we
had a theory that fixed the $a_i$ to automatically give a small value of
$am_{3/2}^2$ in Eq.~(\ref{eq:fake}). Such a theory would then fix the
SSB parameters so that the large logarithms automatically cancel in the
way that we have already mentioned above \cite{am}.\footnote{ Indeed it is
not necessary for all the SSB parameters to be correlated ($n_{\rm
SSB}=1$) in order to obtain $\delbg \simeq \delew$, since many of these
have only a weak effect on $M_Z^2$ in Eq.~(\ref{eq:mZsSig}).}  Needless
to say, we do not have such a theory.  We stress though that this
difference of interpretation of the meaning of $\delew$ is unimportant
for many practical purposes, in particular for providing motivation for
the examination of the phenomenology of SUSY models with low
$\delew$. It also does not detract the importance of $\delew$.

\section{Phenomenology}
\label{sec:phen}

We have argued that 100-300~GeV charged and neutral higgsinos, with a
mass gap of 10-30 GeV with the LSP, characterize natural SUSY scenarios
with $\delew \lesssim 30$.  In this section, we present an overview of
various SUSY signals in such scenarios, with emphasis on signatures 
suggestive of  light higgsinos in the spectrum.

\subsection{LHC} 
\label{subsec:lhc}

\begin{figure}[tbh]{\begin{center}
\includegraphics[width=12cm,clip]{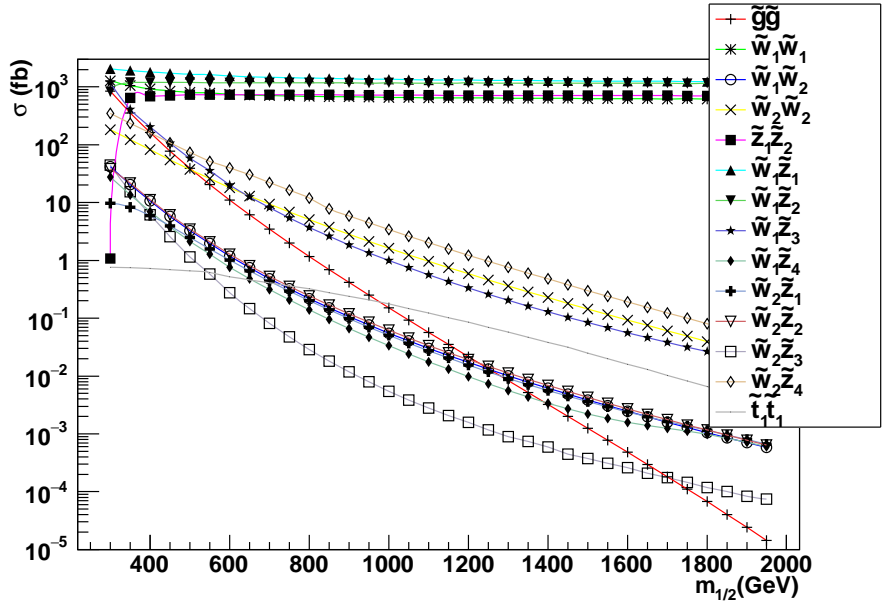}
\caption{Plot of various NLO sparticle pair production cross sections
versus $m_{1/2}$ along the RNS model line (\ref{eq:mline}) for $pp$
collisions at $\sqrt{s}=14$~TeV.  }
\label{fig:csec}\end{center}}
\end{figure}

In natural SUSY the light higgsinos are likely to be the most
copiously produced superpartners at the LHC \cite{rnslhc}. This is
illustrated in Fig.~\ref{fig:csec} where we show various -ino production
cross sections versus $m_{1/2}$, for the RNS model-line with
\be
m_0=5~{\rm TeV}, A_0=-1.6m_0, \tan\beta=15, \mu=150~{\rm GeV}, \ {\rm
  and} \  m_A=1~{\rm TeV},
\label{eq:mline}
\ee
at LHC14. The cross sections for the production of higgsino-like
charginos and neutralinos ($\tw_1$, $\tz_{1,2}$) whose masses are
$\sim |\mu| = 150$~GeV across most of the plot remain flat, while
cross sections for the gaugino-like states ($\tw_2$, $\tz_{3,4}$) fall off
because their masses increase with $m_{1/2}$. Cross sections for associated
gaugino-higgsino pair production are dynamically suppressed.
Despite the sizeable production rate, the small energy release in
their decays makes signals from higgsino pair production difficult to
detect over SM backgrounds. We are thus led to investigate other
strategies for discovery of SUSY.

\subsubsection{Gluinos} 

Gluino pair production leads to the well-known cascade decay signatures
in the widely explored multi-jet + multilepton channels. That lighter
charginos and neutralinos are higgsino-like rather than gaugino-like
affects the relative rates for the various topologies with specific
lepton multiplicity, relative to expectations in mSUGRA. However, the
gluino mass reach which is mostly determined by the gluino production
cross section (for very heavy squarks, the gluino pair production rate is
essentially determined by $m_{\tg}$) and is not significantly altered. An
examination of the gluino reach within the RNS framework shows that
experiments at LHC14 should be sensitive to $m_{\tg}$ values up to
1700~GeV (1900~GeV), assuming an integrated luminosity of 300
(1000)~fb$^{-1}$. It may also be possible to extract the neutralino mass
gap, $m_{\tz_2}-m_{\tz_1}$, from the end-point of the mass distribution
of opposite sign/same flavour dileptons from the leptonic decays of
$\tz_2$ produced in gluino decay cascades, if the mass $\tz_2-\tz_1$
mass gap is large enough \cite{rnslhc}. We note, however, that
experiments at the LHC can discover gluinos only over part of the range
allowed by naturalness considerations.

\subsubsection{Same Sign Dibosons} 

In a typical scenario based on naturalness consideratations we
expect that $|\mu| \ll M_{1,2}$ so that $\tw_1$ and $\tz_2$ are
higgsino-like and only 10-30~GeV heavier than $\tz_1$, $\tz_3$ is
dominantly a bino, and $\tw_2$ and $\tz_4$ are winos. For heavy squarks,
electroweak production of the bino-like $\tz_3$ is dynamically
suppressed since gauge invariance precludes a coupling of the bino to
the $W$ and $Z$ bosons. However, winos have large ``weak iso-vector''
couplings to the vector bosons so that wino production cross sections
can be substantial. Indeed we see from Fig.~\ref{fig:csec} that for high
values of $m_{1/2}$ the kinematically disfavoured $\tw_2^\pm\tw_2^\mp$
and $\tw_2\tz_4$ processes are the dominant sparticle production
mechanisms with large visible energy release and high
$\eslt$.\footnote{The $\tw_1\tz_3$ cross section is also significant,
but falls more steeply with $m_{1/2}$ because the gaugino-higgsino
mixing becomes increasingly suppressed.} Wino production
 leads to a novel signature involving same-sign dibosons
\begin{figure}[tbh]{\begin{center}
\includegraphics[width=10cm,clip]{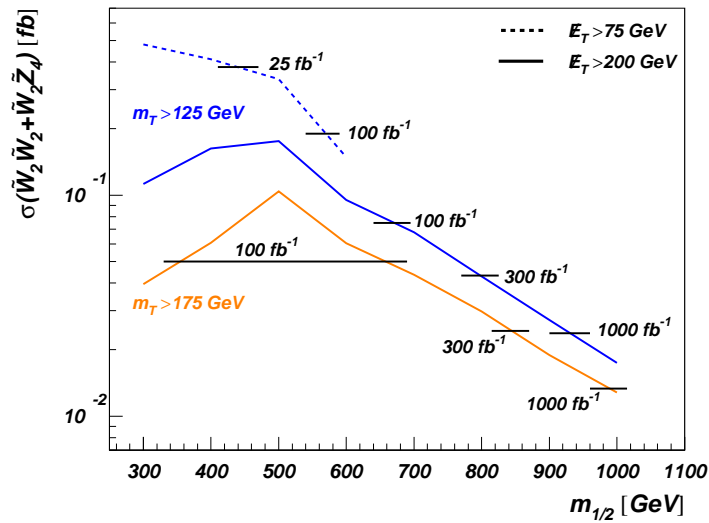}
\caption{Same-sign dilepton cross sections (in {\it fb}) at LHC14 after
cuts vs. $m_{1/2}$ along the RNS model line (\ref{eq:mline}) from
$\tw_2^\pm\tz_4$ and $\tw_2^\pm\tw_2^\mp$ production and calculated
reach for 100, 300 and 1000~fb$^{-1}$.  The upper solid and dashed
(blue) curves require $m_T^{\rm min} >125$~GeV while the lower solid
(orange) curve requires $m_T^{\rm min} >175$~GeV. The dashed and solid 
curves require $\eslt > 75$ or 200~GeV, respectively. The signal is
detectable above the horizontal lines.  }
\label{fig:ssdbreach}\end{center}}
\end{figure}
produced via the process, $pp \to \tw_2^{\pm} (\to
W^{\pm}\tz_{1,2})+\tz_4 (\to W^\pm\tw_1^\mp)$. The decay products of
$\tw_1$ and $\tz_2$ tend to be soft, so that the signal of interest is a
pair of same sign, high $p_T$ leptons from the decays of the $W$-bosons,
{\em with limited jet activity in the event.} This latter feature serves
to distinguish the signal from wino pair production from same sign
dilepton events that might arise at the LHC from gluino pair production
\cite{glssl}.  We note also that $pp \to \tw_2^\pm \tw_2^\mp$ production
(where one chargino decays to $W$ and the other to a $Z$) also makes a
non-negligible contribution to the $\ell^\pm\ell^\pm +\eslt$ channel
when the third lepton fails to be detected. The same sign dilepton
signal with limited jet activity is a hallmark of all low $\mu$ models,
as long as wino pair production occurs at substantial rates.

The extraction of the same sign dilepton signal from wino production
requires a detailed analysis 
to separate the signal from SM backgrounds: see Sec.~5 of
Ref.~\cite{rnslhc}. The most important cuts necessary
for suppressing backgrounds are a hard cut on $\eslt$, together with
a cut on 
\[
m_T^{\rm min} \equiv {\rm
  min}\left[m_T(\ell_1,\eslt), m_T(\ell_2,\eslt)\right].
\] 
The $5\sigma$  reach of the LHC for the NUHM2 model line (\ref{eq:mline}),
chosen to yield low $\delew$, is illustrated in
Fig.~\ref{fig:ssdbreach} as a function of the gaugino mass parameter
$m_{1/2}$.
We show results for relatively soft cuts (dashed line) and hard cuts
(solid lines) on $\eslt$ and $m_T^{\rm min}$. We see that with
300~fb$^{-1}$ (1000~fb$^{-1}$) of integrated luminosity, experiments at
the LHC will probe $m_{1/2}$ values up to 840~GeV (1~TeV), well in
excess of what can be probed via cascade decays of gluinos.

\subsubsection{Hard Trileptons} 

It is natural to examine the SUSY reach via the trilepton channel from
wino pair production, {\it i.e.} from the reaction $pp \to \tw_2 \tz_4 +
X \to W+Z +\eslt+ X$, long considered to be the golden mode for SUSY
searches~\cite{trilep}. Here the $\eslt$ arises from the
$\tw_1/\tz_{1,2}$ (whose visible decay products are very soft) daughters
of the winos. A detailed analysis \cite{rnslhc} shows that the LHC14
reach extends to $m_{1/2} = 500$ (630)~GeV for an integrated luminosity
of 300 (1000)~fb$^{-1}$. This is considerably lower than the reach via
the SSdB channel.

\subsubsection{Four Lepton Signals} 

Light higgsino models also offer the opportunity for detecting SUSY
via $ZZ+\eslt$ events from $\tw_2^+\tw_2^-$ or $\tw_2^\pm \tz_4$
production, when both winos decay to a $Z$ boson  plus a light
chargino/neutralino. This leads to the possibility of a 4 lepton
signal at LHC13. The reach in this channel was also mapped out
for LHC14 in Ref.~\cite{rnslhc}, by requiring 4 isolated leptons with
$p_T(\ell) > 10$~GeV, a $b$-jet veto (to reduce backgrounds from top
quarks), and $\eslt > \eslt({\rm cut})$. The value of $\eslt({\rm
  cut})$ was adjusted to optimize the signal relative to SM
backgrounds from $ZZ, t\bar{t}Z, ZWW, ZZW, ZZZ$ and $Zh(\to WW^*)$
production. Since the background also includes a $Z$ boson, and also
because one of the four leptons in the signal occasionally arises as a
leptonic daughter of the lighter $\tw_1$ or $\tz_2$, requiring lepton
pairs to reconstruct the $Z$-boson mass actually reduces the
significance of the signal. It was found that in low $|\mu|$ models,
the LHC14 reach via the $4\ell$ search extends somewhat beyond that in
the trilepton channel. Indeed an observation of a signal in this
channel together with the SSdB signal would point to a SUSY scenario
with small value of $|\mu|$ and a comparatively larger wino mass,
typical of the RNS framework.

\subsubsection{Soft Trileptons} 

From Fig.~\ref{fig:csec}, we recall that higgsino pair production is the
dominant sparticle production mechanism at the LHC. This leads us to
examine whether the $e\mu\mu$ signal from $\tw_1\tz_2$ might be
observable, since LHC experiments can detect muons with $p_T(\mu)$ as
small as 5~GeV. With this in mind, we examined the shape of the
invariant mass distribution of dimuons in the reaction $pp \to \tw_1(\to
e\nu\tz_1)+\tz_2(\to \mu^+\mu^-\tz_1)$ incorporating cuts to enhance the
soft trilepton signal over large SM backgrounds \cite{rnslhc}. The
signal dimuons, of course, all have a mass smaller than the kinematic
end point at $m_{\tz_2}-m_{\tz_1}$, while the background distribution
extends over a much broader range. It was found that there should indeed
be an enhancement of this distribution at low values of $m(\mu^+\mu^-)$,
so that a {\em shape analysis} of the dimuon mass spectrum may well
reveal the signal if $m_{1/2} < 400-500$~GeV, for
$\mu=150$~GeV.\footnote{This range of $m_{1/2}$ is nearly excluded by
lower limits on the gluino mass \cite{lhcgl}, assuming that gaugino mass
parameters unify.} For larger values of $m_{1/2}$ the mass gap becomes
so small that the resulting spectral distortion is confined to just one
or two low mass bins. In our view, the soft-trilepton signal is unlikely
to be a discovery channel, though it could serve to strikingly confirm a
SUSY signal in the SSdB or multilepton channels. Perhaps more
importantly, an $m_{\mu\mu}$ spectrum distortion in $e\mu\mu+\eslt$
events would point to a small value of $|\mu|$, if model parameters
happen to lie in a fortuitous mass range.

\subsubsection{Mono-jet and Mono-photon Signals} \label{subsec:mono}

Many groups have suggested that experiments at the LHC may be able to
identify the pair production of LSPs via high $E_T$ mono-jet or
mono-photon plus $\eslt$ events, where the jet or the photon arises
from QCD or QED radiation. Many of these studies have been performed
using non-renormalizable contact operators for LSP production
\cite{monojet,mono-ex,monoother}. This grossly overestimates the rates for
mono-jet/mono-photon production at high $E_T$, especially in models
such as RNS where $s$-channel $Z$ exchange dominates LSP pair
production \cite{buch}. A careful study of this signal for the case of
light higgsinos, incorporating the correct matrix elements for all
relevant higgsino pair production processes 
within the RNS framework, shows that it will be very difficult to
extract the signal unless SM backgrounds can be controlled at the
better than the percent level \cite{mono}. The problem is that the
jet/photon $E_T$ distribution as well as the $\eslt$ distribution has
essentially the same shape for the signal and the background.

In Ref.~\cite{ghww} it was suggested that it may be possible to enhance
the mono-jet signal relative to background by requiring additional soft
leptons in events triggered by a hard mono-jet. Ref.~\cite{kribs}
examined the mono-jet signal requiring, in addition, two opposite-sign
leptons in each event, and showed that the SUSY signal could  indeed be 
observable at the LHC.  A subsequent detailed study of mono-jet events with
opposite-sign, same-flavour dileptons with low invariant mass
showed that experiments at LHC14 would be able to detect a $5\sigma$
signal from higgsino pair production for $|\mu|< 170$~(200)~GeV,
assuming an integrated luminosity of 300 (1000)~fb$^{-1}$
\cite{dilep}. We conclude that while LHC experiments will be sensitive
to the most promising part of the parameter of natural SUSY models, they
would not be able to probe the entire RNS region with $\delew \leq 30$.

\subsubsection{Same Sign Charginos from Vector Boson Fusion}

The ATLAS and the CMS experiments have reported a measurement of the
cross section for same-sign $W$ pair production via the ``vector boson
fusion'' process $qq \to q'q'W^{\pm}W^{\pm}$ at the LHC
\cite{lhcvbf}. This process leads to events with two high rapidity jets
in opposite hemispheres, together with a pair of (leptonically-decaying)
same sign $W$-bosons. The observed rate is compatible with SM
expectations.  Motivated by this observation together with the fact that
chargino masses are expected to be close to $M_Z$ in natural SUSY
models, we were led to examine same-sign chargino production from vector
boson fusion.\footnote{Superpartner production by vector boson fusion
has been suggested by several authors going back nearly a decade
\cite{tilman}, and has received recent attention in
Ref.~\cite{ghww,aggies}. Since $\tw_1^+\tw_1^-$ and $\tz_i\tz_j$
production in association with high rapidity jets also occurs by
conventional $q\bar{q}$ initiated processes, we have confined our
attention to same-sign chargino pair production which (for heavy
squarks) dominantly occurs via vector boson fusion: contributions from
$s$-channel processes $W^{\pm*}\to \tw^{\pm}\tw^{\pm}(W^{*\mp}\to
q'\bar{q})$ are suppressed, and also do not lead to hemispherically
separated jets. Our results for same-sign chargino production are in
agreement with Ref.~\cite{ghww} but at variance with those in
Ref.~\cite{tilman}. We also differ by a factor of about 20 with
the topmost curve in Fig.~2
of the second paper of Ref.~\cite{aggies}.  We have contacted these
groups and they have since confirmed that they agree with our
calculation. We thank B. Dutta, T.~Ghosh, A.~Gurrola, T.~Kamon and
especially T.~Plenh and S.~Wu for extensive communications and
discussion about this discrepancy.} We examined the cross section along
the RNS model line (\ref{eq:mline}) for which the lighter chargino mass
remains close to 150~GeV. To our surprise, we found that the
cross section for $pp \to \tw_1^{\pm}\tw_1^{\pm}jj$ evaluated for the
RNS model line (\ref{eq:mline}) falls rapidly from $\sim 0.1$~fb for
$m_{1/2} \sim 200$~GeV (already excluded by LHC gluino searches
\cite{lhcgl}) to $< 0.01$~fb for $m_{1/2} > 800$~GeV \cite{stengel}. We
stress that the sharp fall-off of the cross section with $m_{1/2}$
cannot be for kinematic reasons since the charged higgsino mass
$m_{\tw_1}$ remains close to 150~GeV.

\begin{figure}[tbh]{\begin{center}
\includegraphics[width=10cm,clip]{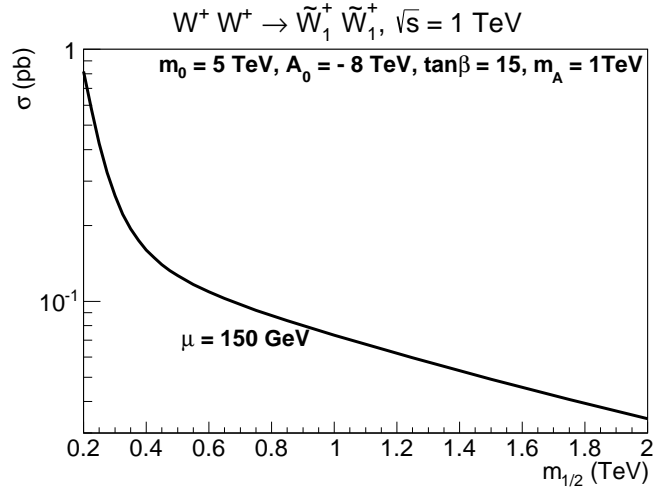}
\caption{Cross section for the reaction $W^+W^+ \to \tw_1^+\tw_1^+$
  vs. $m_{1/2}$ for the RNS model line in Eq.~(\ref{eq:mline}). Note
  that over almost the entire range, $m_{\tw_1}$ is close to 150~GeV.}
\label{fig:WW}\end{center}}
\end{figure}

To better understand the rapid fall-off of the production rate for
same sign chargino pairs, we show the cross section for the underlying
sub-process $W^+W^+ \to \tw_1^+\tw_1^+$ for the same model line in
(\ref{eq:mline}) in Fig.~\ref{fig:WW},  taking
$\sqrt{s}=1$~TeV.\footnote{Since we take squarks to be heavy, $2\to 4$
  amplitudes proportional to $g_s^2g^2$ are suppressed.} We see from
the figure
that this  cross section continues to drop off 
with increasing $m_{1/2}$ {\em even though the lighter chargino mass is
  essentially unchanged over the entire plot.} 

To explain the suppression, we first note that for $m_{1/2} \gg |\mu|$
(or more generally, $M_1, M_2 \gg |\mu|$), there are two Majorana neutralinos
 and a chargino, all with a mass $\simeq |\mu|$, while the gauginos are
 essentially decoupled. In the limit $M_{1,2} \to\infty$, the two
 degenerate neutralinos can be combined into a single Dirac neutralino,
 $\tz_D$, with couplings to the $W\tw_1$ system given by, 
\be
{\cal L} = -|\mu| (\overline{\tw_1}{\tw_1}+\overline{\tz_D}\tz_D) +
\left[\frac{g}{\sqrt{2}}(-i)^{\theta_\mu+1}\overline{\tw_1}\gamma^\mu\tz_D
W_\mu+h.c.\right]
\label{eq:lag}
\ee
We see that the Lagrangian in Eq.~(\ref{eq:lag}) conserves ino-number,
defined to be +1 for the Dirac particles $\tw_1$ and $\tz_D$, -1 for the
corresponding anti-particles, and 0 for sfermions and all SM
particles. Ino-number conservation then requires the cross section for
the process $W^+W^+ \to\tw_1^+\tw_1^+$ must vanish in the limit
$M_{1,2}\to \infty$.\footnote{It is, of course, not possible to
  assign a definite non-zero $U(1)$-charge such as ino-number to a Majorana
  field. The reader can easily verify that the Majorana nature of the
  neutralino is critical for obtaining a non-vanishing amplitude for the
process $W^+W^+\to \tw_1^+\tw_1^+$.}  The couplings of the higgsinos to the
fermion-sfermion system violate ino-number conservation, but the
corresponding amplitudes are very small because we have taken the squark
masses around 5~TeV. We thus understand why the cross section in
Fig.~\ref{fig:WW} is strongly suppressed for $m_{1/2} \gg |\mu|$, and
are forced to conclude that {\em same-sign chargino production is not a
  viable avenue for searching for the light higgsinos of natural
  SUSY} \cite{stengel}.

\subsubsection{Recap of the LHC14 Reach in the RNS Framework}

Table~\ref{tab:reach} summarizes the projected reach of LHC14 in terms 
of the gluino mass within the RNS framework that we advocate  for
phenomenological analyses of natural SUSY.
\begin{table}
\begin{center}
\begin{tabular}{|l|r|r|r|r|}
\hline
 Int. lum. (fb$^{-1}$) & $\tg\tg$ &  SSdB & $WZ\to 3\ell$ &$4\ell$ \\
\hline
\hline
10   & 1.4 &   --  & -- & --\\
100  & 1.6 &  1.6 & -- & $\sim 1.2$\\
300  & 1.7 &  2.1 & 1.4& $\gtrsim 1.4$ \\
1000 & 1.9 &  2.4 & 1.6& $\gtrsim 1.6$ \\
\hline
\end{tabular}
\caption{Reach of LHC14 for SUSY in terms of gluino mass, $m_{\tg}$
  (TeV) for various values of integrated luminosity values along an RNS
  model line introduced in (\ref{eq:mline}).
\label{tab:reach}}
\end{center}
\end{table}
We see that for an integrated luminosity in excess of $\sim
100$~fb$^{-1}$ the greatest reach is attained via the SSdB channel, if
we assume gaugino mass unification. More importantly, the SSdB channel
provides a novel signature for a SUSY signal in {\em any natural model
  of supersymmetry} without a proliferation of (beyond MSSM) particles
at the weak scale. In this case, there may be striking confirmatory
signals in the $4\ell$ and soft-trilepton channels in addition to the
much-discussed clean trilepton signal from wino pair production.

\subsection{Electron-Positron Colliders} 

\begin{figure}[tbh]{\begin{center}
\includegraphics[width=12cm,clip]{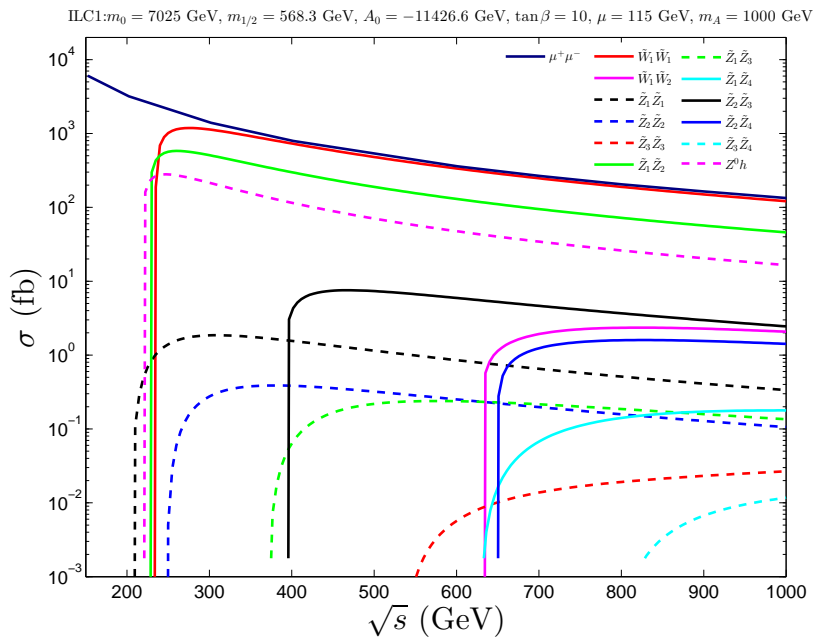}
\caption{Sparticle pair production cross sections as a function of
  centre-of-mass energy for unpolarized beams at an electro-positron
  collider for the low $\delew$ NUHM2 parameter point with parameters
  listed in the figure.}
\label{fig:ee}\end{center}}
\end{figure}

Since light higgsinos are $SU(2)$ doublets, they necessarily have
sizeable electroweak couplings, and so {\em must be} copiously produced
at $e^+e^-$ colliders, unless their production is kinematically
suppressed. This can be seen from Fig.~\ref{fig:ee} where we illustrate
the variation of various sparticle production cross sections at an
electron-positron collider with the centre-of-mass energy $\sqrt{s}$,
for the NUHM2 point with $m_0=7025$~GeV, $m_{1/2}=568.3$~GeV,
$A_0=-11424$~GeV, $\tan\beta=10$, $\mu=115$~GeV and $m_A= 1$~TeV. Indeed
we see that the cross sections for higgsino pair production proceses are
comparable to the cross section for muon pair production if higgsino
production is not
kinematically suppressed. Moreover, the higgsino pair production rate,
for higgsinos with masses comparable to $m_h$ exceeds that for $Zh$
production, so that these facilities may well be higgsino factories in
addition to being Higgs boson factories.
Electron-positron linear colliders
that are being envisioned for construction are thus the obvious
facility for definitive searches for natural SUSY.  The real question
is whether, in light of the small visible energy release in higgsino
decays,it is possible to extract the higgsino signal above SM
backgrounds. These dominantly come from two-photon-initiated processes
because those $2\to 2$ SM reactions can be efficiently suppressed by a cut
on the visible energy in the event.

The higgsino signal was examined in Ref.~\cite{ilc} where the authors
studied two cases. For Case A (which is just the NUHM2 point
shown in Fig.~\ref{fig:ee}), $m_{\tw_1}=117.3$~GeV, $m_{\tz_2}=124$~GeV
and $m_{\tz_1}=102.7$~GeV, with $\delew=14$, and a neutralino mass gap
of 21~GeV. Case B was chosen so that $m_{\tw_1} \simeq
m_{\tz_2}=158$~GeV, and a mass gap with the neutralino of just $\sim
10$~GeV. This case has $\delew =28.5$, close to what we consider the
maximum for natural models, and a neutralino mass gap that is nearly as
small as it can be, consistent with $\delew^{-1}\leq 3\%$. The small
mass gap severely limits the visible energy, and in this sense Case B
represents the maximally challenging situation within the RNS framework.

The most promising signals come from $e^+e^-\to\tw_1(\to
\ell\nu\tz_1)\tw_1(\to q\bar{q}\tz_1)$ which leads to $n_{\ell}=1$,
$n_j=1$ or 2 plus $\eslt$ events, and from $e^+e^- \to \tz_1\tz_2(\to
\ell\ell\tz_1)$ (with 90\% electron beam polarization to reduce $WW$
background) processes.  SM backgrounds can be nearly eliminated using
judicious cuts on the visible energy (signal events are very soft),
$\eslt$ and transverse plane opening angles between leptons and/or
jets. The signal is observable at the 5$\sigma$ level assuming
$\sqrt{s}=250$~GeV (this is the energy for the initial phase of the
linear collider that is being envisioned for construction in Japan) for
Case~A, and $\sqrt{s}=340$~GeV for Case B, with an integrated luminosity
of just a few fb$^{-1}$. We refer the reader to Ref.~\cite{ilc} for
details. Based on this study, we infer that an electron-positron
collider will be able to detect higgsino-pair production nearly all the
way to the kinematic limit provided that the neutralino mass gap is not
smaller than $\sim 10$~GeV, and further, that an electron-positron
collider with $\sqrt{s}=600$~GeV will probe the entire parameter space
with $\delew \leq 30$ in the RNS framework.

Aside from discovery, the clean environment of electron-positron
collisions also enables precise mass measurements. For example, even in
the maximally difficult Case B considered in Ref.~\cite{ilc}, a fit to
the invariant mass distribution of dileptons in $\tz_1\tz_2$ events
allows the determination of the neutralino mass gap,
$m_{\tz_2}-m_{\tz_1} =9.7\pm 0.2$~GeV. A subsequent fit to the
distribution of the total energy of the two leptons then allows the
extraction of individual neutralino masses: $m_{\tz_2}=158.5 \pm
0.4$~GeV and $m_{\tz_1}= 148.8 \pm 0.5$~GeV. These mass determinations,
together with cross section measurements using polarized beams, point to the
production of higgsinos as the underlying new physics, and suggest a
link to a natural origin of gauge and Higgs boson masses \cite{ilc}.

\subsection{Precision Measurements}

Generally speaking, precision measurements of SM particle properties
offer an independent avenue (from direct production of new particles)
for discovery of new physics. This is not, however, the case for the RNS
scenario where our assumption that GUT scale matter sfermion masses are
very large essentially precludes any observable effect. (We emphasize
that this choice was not required by naturalness considerations, but
made to alleviate issues with flavour physics.) For instance, the RNS
contributions to the rate for the inclusive $b\to s\gamma$ decay are
very suppressed. This is in keeping with the fact that the SM prediction
\cite{bsgsm} for the branching ratio $B(b\to s\gamma)$ is compatible,
within errors, with its measured value \cite{bsgexpt}. Likewise, it is
not possible to attribute the reported deviation of the measured value
of the muon anomalous magnetic moment \cite{muonmeas} from its
expectation in the SM \cite{muontheory} to SUSY contributions within the
RNS framework. New physics beyond the RNS framework will be needed to
account for this discrepancy, if the SM computation of $(g_\mu-2)$ holds
up to scrutiny. Finally, we note that though SUSY contributions to the
rare decay rate for the exclusive decay $B_s\to \mu^+\mu^-$ do not
decouple with the super-partner mass scale, these are strongly
suppressed for large values of $m_A$. Recall that for moderate to large
values of $\tan\beta$, $m_A^2 \simeq m_{H_d}^2 - m_{H_u}^2$ aside from
radiative corrections. Thus, for large values of $m_{H_d}^2$, $m_A^2$
can easily be in the multi-TeV range without jeopardizing electroweak
fine-tuning because the contribution of the $m_{H_d}^2$ term in
Eq.~(\ref{eq:mZsSig}) is suppressed by the $(\tan^2\beta-1)$
factor. This is fortunate because the measured value \cite{bsmmmeas} for
the branching fraction for this process is also in good agreement with
the SM prediction \cite{bsmmtheory} so that any new physics contribution
is strongly constrained.

\subsection{Dark Matter} 

Since the LSP is likely higgsino-like in all simple models with natural
supersymmetry, it will annihilate rapidly to gauge bosons (via its large
coupling to the $Z$ boson, and also via $t$-channel higgsino exchange
processes) in the early universe. Thus, in natural supersymmetry the
measured cold dark matter density {\em cannot arise solely from
thermally produced higgsinos} in standard Big Bang cosmology. Dark
matter is thus likely to be multi-component. It is important to note
that because naturalness considerations also impose an upper bound on
$m_{\tg}$ and corresponding limits on electroweak gaugino masses (via
gaugino mass unification), {\em the thermal higgsino relic density
cannot be arbitrarily small.}  Indeed, within the RNS framework,
$\Omega_{\tz_1} h^2$ must be between $\sim 0.004-0.03$, as shown by
Baer, Barger and Mickleson \cite{bbm}. This has implications for DM
detection experiments. Specifically, ton-size direct detection
experiments such as Xe-1~Ton that are sensitive to a spin-independent
nucleon-LSP cross section at the $10^{-47}$~pb level will be able to
detect a signal over the entire range of RNS parameters with $\delew
\leq 30$.\footnote{We remind the reader that there are the usual caveats
to this conclusion. For instance, if physics in the sector that makes up
the remainder of the dark matter entails late decays that produce SM
particles, the neutralino relic density today could be further diluted,
reducing the signal; see {\it e.g.} Ref.~\cite{howie_axion}. On the
other hand, late decays of associated saxion, axino or even
string-moduli fields to the neutralino could enhance the neutralino
relic density from its thermal value. The important lesson is that while
the thermal relic density is interesting to examine, it would be
imprudent to categorically exclude a new physics scenario based on relic
density considerations alone, because the predicted relic density can be
altered by the unknown (and, perhaps, unknowable) history of the Early
Universe \cite{graciela}.}  Thus, the outcome of these experiments has
an important impact on naturalness.

\subsection{Non-Universal Gaugino Masses}

The RNS framework {\em assumes} gaugino mass universality. It is,
however, possible that the bino and wino mass parameters are
independent of the gluino mass and fotuitously small. This does not
have any impact on $\delew$ but will affect both collider as well as
dark matter phenomenology \cite{nonuni}. In particular, if binos
and/or winos are accessible at the LHC (with $|\mu|$ also small for
naturalness reasons), signals from $\tz_{3,4}$ as well as $\tw_2$
production at the LHC would occur at observable rates and be
relatively straightforward
to detect  because the mass
difference between these states and the higgsinos is typically
large. Multilepton+$\eslt$ events, $WZ+\eslt$ events \cite{wz} and $Wh+\eslt$
events \cite{wh} would be typical in such scenarios. Experiments at
the LHC are already searching for these signals \cite{lhcwzwh}. The dark
matter could be a well-tempered neutralino which saturates the dark
matter relic density if the bino is light, but would necessarily have
to have other components (the axion and its associated SUSY partners,
or hidden sector particles are obvious candidates) if instead $M_2$ is small.

\section{Concluding Remarks} 
\label{sec:concl}

Weak scale supersymmetry stabilizes the electroweak scale and, in our
view, offers the best solution to the big hierarchy problem. The
non-observation of superpartners at LHC8 have led some authors \cite{lykken} to
express reservations about this far-reaching idea. As far as we can ascertain,
these are largely based on the
early notions of fine-tuning that ignore the possibility that the
underlying soft-SUSY-breaking parameters of the underlying theory might
be correlated. While we acknowledge that a credible high scale model of
SUSY breaking that {\em predicts} appropriate correlations among the SSB
parameters
and so
automatically has a modest
degree of fine-tuning has not yet emerged, obituaries of supersymmetry
when the LHC has run at just 60\% of its design energy and collected
$\lesssim 10\%$ of the anticipated integrated luminosity seem to be premature. 

SUSY GUT ideas pioneered by Arnowitt, Chameseddine and Nath
\cite{acnsugra} and others in the 1980s remain as promising as
ever. Moreover, the original aspirations of early workers on weak scale
supersymmetry outlined in \ref{subsec:aspir} remain unchanged, if we
accept that: \bi
\item ``accidental cancellations'' at the
few percent level are ubiquitous and may not require explanation, and
\item dark matter may be multi-component.\footnote{Given that visible
  matter which comprises a small mass fraction of the total matter content 
  already consists of several components, this is hardly a stretch.}
\ei

Viable natural {\em spectra} with light higgsinos exist without a need
for weak scale new particles beyond the MSSM. We have argued that light
higgsinos are necessary at least for the most economic realizations of
the ideas of SUSY naturalness (see, however, the qualifying discussion
in Sec.~\ref{subsec:ewft}), and {\em may} yield novel signals
for supersymmetry at the LHC. We have analysed these within the RNS
framework which has a low value of $\delew$ by construction, and so
leads to a SUSY spectrum that could have its origin in an underlying
natural theory. Since many phenomenological
results are sensitive to just the sparticle spectrum, these can
be abstracted using the RNS framework which we view as a proxy (for
phenomenological purposes) for the underlying natural SUSY theory.

\begin{figure}[tbh]{\begin{center}
\includegraphics[width=10cm,clip]{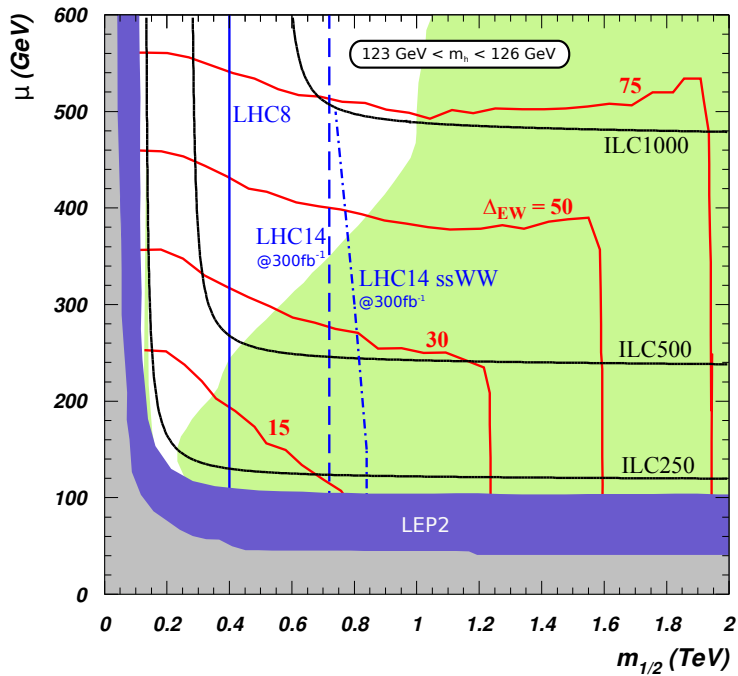}
\caption{Plot of $\delew$ contours (red) labelled by the value of
$\delew=15,30,50$ and 75 in the $m_{1/2}\ vs.\ \mu$
plane of NUHM2 model for $A_0=-1.6 m_0$ and $m_0=5$~TeV and $\tan\beta
=15$.  We show the region accessible to LHC8 gluino pair searches
(solid blue contour), and the region accessible to LHC14 searches with
300~fb$^{-1}$ of integrated luminosity (dashed and dot-dashed contours).
LHC14 experiments will also be sensitive to a mono-jet plus opposite-sign, 
same-flavour, low mass dilepton signal if
$\mu \lesssim 170-200$~GeV.
We also show the reach of various $e^+e^-$  colliders for higgsino pair
production (black contours).  The very light-shaded (green) region has
$\Omega_{\tz_1}^{std}h^2<0.12$.  The dark  (light) shaded region along
the axes is
excluded by LEP2 (LEP1) searches for chargino pair production.  To aid
the reader, we note that $m_{\tg}\simeq 2.5 m_{1/2}$.}
\label{fig:pan}\end{center}}
\end{figure}

RNS phenomenology is discussed in Sec.~\ref{sec:phen} and summarized in
Fig.~\ref{fig:pan}, where we show contours of $\delew$ in the
$m_{1/2}-\mu$ plane of the NUHM2 model with large $m_0$ and $A_0=-1.6m_0$.
Above and to the right of the $\delew=30$ contour, we regard the
spectrum to be fine-tuned: in this region the fine-tuning {\em must be}
worse than $\delew^{-1} \sim 3$\%. The light-shaded (green) region is
where the thermal higgsino relic density is smaller than its measured
value, with the balance being made up by something else. It is worth
stressing that despite the fact that thermal relic higgsinos of RNS
comprise only a fraction of the dark matter, ton scale direct detection
experiments will be able to detect the higgsino signal.  The dashed line
shows the LHC14 reach via the canonical search for gluinos, while the
dot-dashed line shows the projected reach via searches in the novel SSdB
channel discussed in Sec.~\ref{subsec:lhc}. The region with $\mu < 170$
(200)~GeV may be probed via searches for hard mono-jet events with low
mass, same-flavour, opposite sign dileptons as discussed in
Sec.~\ref{subsec:mono}. We see, however, that LHC searches will, by
themselves, not be able to cover the entire parameter space with $\delew
< 30$. The remainder of this parameter space should be accessible, via a
search for higgsinos at an $e^+e^-$ collider operating at
$\sqrt{s}=600$~GeV. Such a facility will be a decisive probe of light
higgsinos associated with a natural origin of Higgs and gauge boson
masses.

In summary, the fact that low scale physics is only logarithmically
(and not quadratically) sensitive to the scale of ultra-violet physics
remains a very attractive feature of softly broken SUSY models that
provides an elegant resolution of the big hierarchy problem.  That it
is possible to find phenomenologically viable models with low
electroweak fine-tuning leads us to speculate that our understanding
of UV physics is incomplete, and that there might be high scale models
with the required parameter correlations that will lead to comparably
low values of the true fine-tuning parameter $\delbg$. The
supergravity GUT paradigm remains very attractive despite the absence
of sparticle signals at LHC8. We remain hopeful that experiments at
the new run of the LHC will unearth new physics and perhaps realize the
vision laid out by ACN and other colleagues during the 1980s.

\section*{Acknowledgments}
I am grateful to H.~Baer, P.~Huang, V.~Barger, D.~Mickelson,
A.~Mustafayev, M.~Padefkke-Kirkland, W.~Sreethawong and P.~Stengel for
discussions and collaboration on much of the work described here. I have
learnt much from the many disagreements we have had, and continue to
have, about the interpretation of fine-tuning.  I thank Howie
Baer, Azar Mustafayev and Pran Nath for comments on the text.  This work
was supported in part by the US Department of Energy.

%


\begin{thebibliography}{99}
%
\bibitem{nath} P. Nath, {\em Phys. Scripta} {\bf 90} (2015) 6, 068007
[arXiv:1502.00639]

%
\bibitem{wssusy} A.~Neveu and J.~Schwarz, {\em Nucl. Phys.} {\bf B31}
  (1971) 86; P.~Ramond, {\em Phys. Rev} {\bf D3} (1971) 2415; J.~Gervais
  and B.~Sakita, {\em Nucl. Phys.} {\bf 34} (1971)  632.
%
\bibitem{golfand} Y.~Golfand and E.~Likhtman, {\em JETP Lett.} {\bf 13}
  (1971) 323.
%
\bibitem{volkov} D.~Volkov and V.~Akulov, {\em JETP Lett.} {\bf 16}
  (1972) 621.
%
\bibitem{wessz} J.~Wess and B.~Zumino, {\em Nucl. Phys.} {\bf B70} (1974)
  39. 


%
\bibitem{fayet} P.  Fayet, {\em Nucl. Phys.} {\bf B90} (1975) 104; 
P. Fayet, {\em Phys. Lett.} {\bf B64} (1976) 159; 
P.~Fayet,
  {\em Phys. Lett.} {\bf B69} (1977) 489;
G.~Farrar and P.~Fayet, {\em  Phys. Lett.} {\bf B76} (1978) 575.


%
\bibitem{hier} E.~Witten, {\em Nucl. Phys.} {\bf B188} (1981) 513;
S.~Dimopoulos and H.~Georgi, {\em Nucl. Phys.} {\bf B193} (1981),150;
N.~Sakai, {\em Z. Phys.} {\bf C11} (1981) 153;  R.~Kaul, {\em Phys. Lett.} {\bf B109} (1982) 19.
%

\bibitem{susswein} E.~Gildener and S.~Weinberg,  {\em Phys. Rev.} {\bf D13} 
(1976) 3333; E.~Gildener, {\em Phys. Rev.} {\bf D14} 
(1976) 1667; L.~Susskind, {\em Phys. Rev.} {\bf D20} (1979) 2619.

%
\bibitem {girardello} L.~Girardello and M.~Grisaru, {\em Nucl. Phys.}
  {\bf B194} (1982) 65.
%

\bibitem{oraif} L.~O'Raifeartaigh, {\em Nucl. Phys.} {\bf B96} (1975)
  331.
%
\bibitem{FI} P.~Fayet, and J.~Illiopoulos, {\em Phys. Lett.} {\bf B51}
  (1974) 461.


%
\bibitem{sumrule} F.~Ferrara, L.~Girardello and F.~Palumbo, {\em
  Phys. Rev.} {\bf D20} (1979) 403.
%
\bibitem{mssm} S.~Dimopoulos and H.~Georgi, Ref.~\cite{hier};
N.~Sakai, Ref.~\cite{hier}.
%
\bibitem{vs} D.~Volkov and V.~Soroka, {\em JETP} {\bf 18} (1973) 312.
%
\bibitem{sugrarev} For reviews of supergravity and detailed references,
  see {\it e.g.} P. van Nieuwenhuizen, {\em Phys. Rep.} {\bf 68} (1981)
  189.
%
\bibitem{hist} For an account of the development of supersymmetric
  theories, see {\em The Supersymmetic World}, G.~Kane and M.~Shifman
  (World Scientific, 2000); see also, K.~Olive, S.~Rudaz and M.~Shifman,
  Editors, {\em Proceedings of the  International Symposium Celebrating 
30 Years of Supersymmetry, Nucl. Phys.} {\bf B} (Proc. Suppl.) 101 (2001). 
%
\bibitem{ansugra} R.~Arnowitt and P.~Nath, {\em Phys. Lett.} {\bf B56}
  (1975) 177, and R.~Arnowitt, P.~Nath and B. Zumino, {\em Phys. Lett.}
  {\bf B56} (1975) 81.


%
\bibitem{cremmer} E.~Cremmer, S.~Ferrara, L.~Girardello and A.~van
  Proeyen, {\em Phys. Lett.} {\bf B116} (1982) 231 and {\em Nucl. Phys.}
  {\bf B212} (1983) 413.
%
\bibitem{cremmerearly} E.~Cremmer, B.~Julia, J.~Scherk, S.~Ferrara, L.~Girardello and P.~van
  Nieuwenhuizen, {\em Phys. Lett.} {\bf B79} (1978) 231 and {\em Nucl. Phys.}
  {\bf B147} (1979) 105.
%
\bibitem{acnbook}  P.~Nath, R.~Arnowitt and A.~Chamseddine,  {\em
  Applied $N=1$ Supergravity},
  Lectures at 1983 Summer Workshop on Particle Physics, NUB-2613. 
%
\bibitem{wss} H.~Baer and X.~Tata, {\em Weak Scale Supersymmetry}
  (Cambridge, 2006).
%
\bibitem{acnsugra} A.~Chamseddine, R.~Arnowitt  and P.~Nath, {\em
  Phys. Rev. Lett.} {\bf 49} (1982) 970.
%
\bibitem{others} R.~Barbieri, S.~Ferrara and
  C.~Savoy, {\em Phys. Lett.} {\bf B119} (1982) 343; 
N.~Ohta,  {\em Prog. Theor. Phys.} {\bf 70} (1983) 542; 
  L.~Hall, J.~Lykken and  S.~Weinberg, {\em Phys. Rev.}{\bf D27} (1983) (2359).
%
\bibitem{nilles} For an early overview of supergravity model building,
  see H-P. Nilles, {\em Phys. Rep.} {\bf
110} (1984) 1.
%
\bibitem{soniweldon} S.~Soni and H.~Weldon, {\em Phys. Lett.} {\bf B126}
  (1983) 215.


%
\bibitem{radewsb}L.~Iba\~nez and G.~Ross, {\em Phys. Lett.} {\bf B110}
  (1982) 215; K.~Inoue, A.~Kakuto, H.~Komatsu and S.~Takeshita, {\it
  Prog. Theor. Phys.} {\bf 68} (1982) 927, and {\bf 71} (1984) 413;
  L.~Iba\~nez, {\em Phys. Lett.}  {\bf B118} (1982) 73; J.~Ellis,
  J.~Hagelin, D.~Nanopoulos and M.~Tamvakis, {\em Phys. Lett.} {\bf
    B125} (1983) 275; L.~Alvarez-Guam\'e, J.~Polchinski and M.~Wise,
  {\em Nucl. Phys.} {\bf B221} (1983) 495.
%
\bibitem{gaugeunif} U.~Amaldi, W.~de Boer and
  H~F\"urstenau, {\em Phys. Lett.} {\bf B260} (1991) 447; J.~Ellis,
  S.~Kelley and D.~Nanopoulos, {\em Phys. Lett.} {\bf B260} (1991) 131; 
P.~Langacker and M.~Luo, {\em Phys. Rev.} {\bf D44} (1991) 871. 
%
\bibitem{drees}  M.~Drees, arXiv: hep-ph/0501106.
%
\bibitem{knpap} See {\it e.g.}
   R.~Kitano and Y.~Nomura, {\em Phys. Lett.} {\bf B631} (2005) 58  and 
  {\em Phys. Rev.} {\bf D73} (2006) 095004;
M.~Papucci, J.~T.~Ruderman and A.~Weiler,
{\em J. High Energy Phys.} {\bf 1209} (2012) 035.
%
%
\bibitem{eenz} J.~R.~Ellis, K.~Enqvist, D.~V.~Nanopoulos and F.~Zwirner,
  {\em Mod.\ Phys.\ Lett.}  {\bf A1} (1986) 57. 
%
\bibitem{bg}  R.~Barbieri and G.~Giudice, {\em Nucl. Phys.} {\bf B306}
  (1988) 63.
%
\bibitem{baersugra}H.~Baer, V.~Barger, P.~Huang, D.~Mickelson,
  A.~Mustafayev and X.~Tata, 
  {\em Phys. Rev.}  {\bf D87} (2013) 035017.
%
\bibitem{rns} H.~Baer, V.~Barger, P.~Huang, D.~Mickelson, A.~Mustafayev
  and X.~Tata, {\em Phys. Rev.} {\bf D87} (2013) 115028.
%
\bibitem{ltr}  H.~Baer, V.~Barger, P.~Huang, A.~Mustafayev and X.~Tata,
 {\em Phys. Rev. Lett.} {\bf 109} (2012) 161802.
%
\bibitem{CCN} K.~Chan, U.~Chattopadhyay and P.~Nath, {\em Phys. Rev.}
{\bf D58} (1998) 096004.
%
\bibitem{gm} G.~Giudice and A.~Masiero, {\em Phys. Lett.} {\bf B206} (1988) 
480.
%
%
\bibitem{nr}A.~Nelson and T. Roy, {\em Phys. Rev. Lett.} {\bf 114}
(2015) 201802.
%
\bibitem{martin} S.~Martin, arXiv:1506.02105 [hep-ph]. 
%
\bibitem{ckl} T.~Cohen, J.~Kearney and M.~Luty, {\em Phys. Rev.} {\bf
  D91} (2015) 075004.

%
\bibitem{perel} M.~Perelstein and C.~Spethmann, {\em J. High Energy
Phys.} {\bf 0704} (2007) 070.
%
\bibitem{reduce} K.~Chan, U.~Chattopadhyay and P.~Nath, Ref.~\cite {CCN};
P.~Chankowski, J.~Ellis, M.~Olechowski and S.~Pokorski, {\em
Nucl. Phys.} {\bf B544} (1999) 39; S.~King and G.~Kane, {\em Phys. Lett.}
{\bf B451} (1999) 113; S.~Antusch, L.~Calibbi, V.~Maurer, M.~Monaco and
M.~Spinrath, {\em Phys. Rev.} {\bf D85} (2012) 035025; S.~Antusch,
L.~Calibbi, V.~Maurer, M.~Monaco and M.~Spinrath, {\em J. High Energy
Phys.} {\bf 1301} (2013) 187.
%
\bibitem{am} A.~Mustafayev and X.~Tata, invited contribution in volume
  commemorating C.V. Raman's 125th birth annivarsary, {\em Indian
  J. Phys.} {\bf 88} (2014) 991.
%
%
\bibitem{nuhm2} D.~Matalliotakis and H. P.~Nilles, {\em Nucl. Phys.} {\bf
    B435} (1995) 115; V.~Berezinsky, A.~Bottino, J.~Ellis, A.~Fornengo,
    G.~Mignola and S.~Scopel, {\em Astropart. Phys.}
    {\bf 5} (1996) 1; P.~Nath and R.~Arnowitt, {\em Phys. Rev.} {\bf D56}
    (1997) 2820; J.~Ellis, K.~Olive and Y.~Santoso,
    {\em Phys. Lett.} {B539} (2002) 107; 
J.~Ellis, T.~Falk, K.~Olive and Y.~Santoso,
    {\em Nucl. Phys.} {\bf B652} (2003) 259; 
  H.~Baer, A.~Mustafayev, S.~Profumo, A.~Belyaev
    and X.~Tata, {\em J. High Energy. Phys.} {\bf 0507} (2005) 065.
%
\bibitem{rnslhc} H.~Baer, V.~Barger, P.~Huang, D.~Mickelson,
  A.~Mustafayev, W.~Sreethawong and X.~Tata {\em J. High Energy Phys.} {\bf
  1312} (2013) 013 and {\em  J. High Energy Phys.} {\bf
  1506} (2015) 053 (Erratum).
%
\bibitem{meas}H.~Baer, V.~Barger and D.~Mickelson, {\em
Phys. Rev.} {\bf D88} (2013) 095013.
%
\bibitem{lhchiggs} G.~Aad {\it et al.} [ATLAS and CMS Collaborations] {\em Phys. Rev. Lett.} {\bf 114} (2015) 191803.
%
\bibitem{intra} H.~Baer, V.~Barger, M.~Padeffke-Kirkland and X.~Tata,
{\em Phys. Rev.} {\bf D89}  (2014) 037701.
%
\bibitem{flavdec} 
M.~Dine, A.~Kagan and S.~Samuel, {\em Phys. Lett.}
  {\bf 243} (1990) 250;
  A.~Cohen, D.~B.~Kaplan and A.~Nelson, {\em Phys. Lett.} {\bf B388}
  (1996) 588; J.~Bagger,
  J.~Feng and N.~Polonsky, {\em Nucl. Phys.} {\bf B563} (1999) 3.

%
\bibitem{proton} H.~Murayama and A. Pierce, {\em Phys. Rev} {\bf D65}
  (2002) 055009.

\bibitem{fp} J.~L.~Feng, K.~T.~Matchev and T.~Moroi,
  {\em Phys. Rev.} {\bf 61} (2000) 075005; 
J.~L.~Feng and K.~T.~Matchev,
{\em Phys. Rev.} {\bf D63} (2001) 095003;
J.~L.~Feng, K.~T.~Matchev and D.~Sanford,
{\em Phys. Rev.} {\bf 85} (2012) 075007;

%
\bibitem{siege} H.~Baer, V.~Barger, D.~Mickelson and
  M.~Padeffke-Kirkland, {\em Phys. Rev.} {\bf D89} (2014)  115019.
%
\bibitem{baervol}H.~Baer, V.~Barger and M.~Savoy, {\em Physica Scripta}
  {\bf 90} (2015) 6, 068003, arXiv:1502.04127.
%
\bibitem{glssl} V.~Barger, W.-Y.~Keung and R.~Phillips, {\em
  Phys. Rev. Lett.} {\bf 55} (1985) 166; H.~Baer, X.~Tata and
  J.~Woodside, {\em Phys. Rev.} {\bf D45} (1992) 142; R.~Barnett, J.~Gunion and
  H.~Haber, {\em Phys. Lett.} {\bf B315} (1993) 349.
\bibitem{trilep} A.~Chamseddine, P.~Nath and R.~Arnowitt, \plb{129}{1983}{445};
D.~Dicus, S.~Nandi and X.~Tata, \plb{129}{1983}{451};
H.~Baer, K.~Hagiwara and X.~Tata, \prl{57}{1986}{294}
and \prd{35}{1987}{1598};
R.~Arnowitt and P.~Nath, {\em Mod. Phys. Lett.} {\bf A2} (1987) 331;
H.~Baer, C.~H.~Chen, F.~Paige and X.~Tata, \prd{53}{1996}{6241};
H.~Baer, T.~Krupovnickas and X.~Tata, \jhep{0307}{2003}{020}.
%
\bibitem{lhcgl} G.~Aad {\it et al.}  [ATLAS Collaboration], {\em J. High
Energy Phys.} {\bf 1409} (2014) 176 and
{\em J. High Energy Phys.} {\bf 1504} (2015) 116;
 S.~Chatrchyan {\it et
  al.}  [CMS Collaboration], {\em J. High Energy Phys.} {\bf 1406}
  (2014) 055; Y.~Khachatryan {\it et al.} [CMS Collaboration] {\em
  J. High Energy Phys.} {\bf 1505} (2015) 078.
%
%
\bibitem{monojet} H.~Zhang, Q.~Cao, C.~Chen and C. Li, {\em J. High
  Energy Phys.} {\bf 1108} (2011) 018; M.~Beltran, D.~Hooper,
  E.~W.~Kolb, Z.~A.~C.~Krusberg and T.~M.~P.~Tait, {\em J. High Energy
  Phys.} {\bf 1009} (2010) 037; J.~Goodman, M.~Ibe, A.~Rajaraman,
  W.~Shepherd, T.~M.~P.~Tait and H.~B.~Yu, {\em Phys. Rev.} {\bf 82}
  (2010) 116010; A.~Rajaraman, W.~Shepherd, T.~M.~P.~Tait and
  A.~M.~Wijangco, {\em Phys. Rev.} {\bf D84} (2011) 095013; P.~J.~Fox,
  R.~Harnik, J.~Kopp and Y.~Tsai, {\em Phys. Rev.} {\bf D85} (2012)
  056011; A.~Anandkrishnan, L.~Carpenter and S.~Raby, {\em Phys. Rev.}
  {\bf 90} (2014) 055004; J.~Bramante, A.~Delgado, F.~Elahi, A.~Martin
  and B.~Ostdiek, {\em Phys. Rev.} {\bf D90} (2014) 095008.
%
%
\bibitem{mono-ex} 
  S.~Chatrchyan {\it et al.}  [CMS Collaboration],
{\em J. High Energy Phys.} {\bf 1209} (2012) 094;
  G.~Aad {\it et al.}  [ATLAS Collaboration],
{\em J. High Energy Phys.} {\bf 1304} (2013) 075;
   V.~Khachatryan {\it et al.}  [CMS Collaboration],
 \epjc{75}{2015}{235};
  S.~Chatrchyan {\it et al.}  [CMS Collaboration],
 {\em Phys. Rev. Lett.} {\bf 108} (2012) 261803;
G.~Aad {\it et al.}  [ATLAS Collaboration],
{\em Phys. Rev. Lett.} {\bf 110} (2013) 011802.
%
\bibitem{monoother} Mono-$W$, mono-$Z$ and mono-Higgs production has
  also been considered; see Ref.~\cite{dilep} below for references to the
  literature.  Mono-stop signals from the $b$-quark content of the
  proton have recently been considered by K.~Hikasa, J.~Li, L.~Wu and
  J.~M.~Yang, arXiv:1505.06006 [hep-ph].
%
\bibitem{buch} O.~Buchmueller, M.~Dolan and C.~McCabe, {\em J. High
Energy Phys.} {\bf 1401} (2014) 025.
%
\bibitem{mono} H.~Baer, A.~Mustafayev and X.~Tata, \prd{89}{2014}{055007};
C.~Han, A.~Kobakhidze, N.~Liu, A.~Saavedra, L.~Wu  and J.~M. Yang,
 {\em J. High
Energy Phys.} {\bf 1402} (2014) 049.
P.~Schwaller and J.~Zurita,  {\em J. High
Energy Phys.} {\bf 1403} (2014) 060; D.~Barducci, A.~Belyaev,
A.~Bharucha, W.~Porod and V.~Sanz, arXiv:1504.02472 express a more
optimistic viewpoint for detection of the monojet signal.
%
\bibitem{ghww} G~Giudice, T.~Han, K.~Wang and L-T.~Wang, {\em
  Phys. Rev.} {\bf D81} (2010) 115011.
%
\bibitem{kribs}Z.~Han, G.~Kribs, A.~Martin and A.~Menon,  {\em
  Phys. Rev.} {\bf D89} (2014) 075007.
%
\bibitem{dilep} H.~Baer, A.~Mustafayev and X.~Tata,  {\em
  Phys. Rev.} {\bf D90} (2014) 115007.
%
\bibitem{lhcvbf} G.~Aad {\it et al.} (ATLAS Collaboration)
{\em Phys. Rev. Lett.} {\bf 113} (2014) 141803;
V.~Khachatryan {\it et al.} (CMS Collaboration)
{\em Phys. Rev. Lett.} {\bf 114} (2015) 051801.

%
\bibitem{tilman} G.~Cho, K.~Hagiwara, T.~Plehn and D.~Rainwater,  {\em
  Phys. Rev.} {\bf D73} (2006) 054002.
%
\bibitem{aggies} A.~Delannoy {\it et al.} {\it Phys. Rev. Lett.} {\bf
  111} (2013) 061801; B.~Dutta {\it et al.} {\em
  Phys. Rev.} {\bf D87} (2013) 035029; B.~Dutta {\it et al.} {\em
  Phys. Rev.} {\bf D90} (2014) 095022; B.~Dutta {\it et al.} {\em
  Phys. Rev.} {\bf D91} (2015) 055025.
%
\bibitem{stengel} P.~Stengel and X.~Tata, paper in preparation.
%
\bibitem{ilc} H.~Baer, V.~Barger, D.~Mickelson, A.~Mustafayev and X.~Tata,
{\em J. High Energy Phys.} {\bf 1406} (2014) 172.
%
\bibitem{bsgsm} M.~Misiak {\it et al.} {\em Phys. Rev. Lett.}  {\bf
  98} (2007) 022002.

\bibitem{bsgexpt} J.~P. Lees {\it et al.} (BaBar Collaboration) {\em
  Phys. Rev.} {\bf D86} (2012) 052012; T.~Saito {\it et al.} (Belle
  Collaboration) {\em Phys. Rev} {\bf D91} (2015) 052004.


\bibitem{muonmeas} G.~Bennett {\it et al.} (E821 Experiment) {\em
  Phys. Rev.} {\bf D73} (2006) 072003.

\bibitem{muontheory} F.~Jegerlehner and A.~Nyffler, {\em Phys. Rep.}
  {\bf 477} (2009) 1; M~Davier {\it et al.} \epjc{71}{2011}{1515};
  K.~Hagiwara, R.~Liao, A.~Martin, D.~Nomura and T.~Teubner, {\em
    J. Phys. G} {\bf 38} (2011) 085003.

\bibitem{bsmmmeas} V.~Khachatryan {\it et al.} (CMS and LHCb Collaborations)
{\em Nature} {\bf 522} (2015) 68.

\bibitem{bsmmtheory} See C.~Bobeth, M.~Gorbahn, T.~Herman, M.~Misiak,
  E.~Stamou and M.~Steinhauser, {\em Phys. Rev. Lett.} {\bf 112}
  (2014) 101801 ofr an updated evaluation. 

%
\bibitem{bbm} H.~Baer, V.~Barger and D.~Mickelson, {\em
Phys. Lett.} {B726} (2013) 330.



%
\bibitem{howie_axion}H.~Baer, A.~Lessa, {\em J. High Energy Phys.} {\bf
  1106} (2011) 027; H.~Baer, A.~Lessa and W.~Sreethawong, {\em JCAP} {\bf
  1106} (2011), 036; see K. J.~Bae, H.~Baer, V.~Barger, M.~Savoy and
  H.~Serce, arXiv:1503.04137 [hep-h] for dark matter contributions from
  the axion sector in the context of the RNS scenario; see K. J.~Bae,
  H.~Baer and Lessa, arXiv:1306.2986 [hep-ph] for an overview.
%
\bibitem{graciela} G.~Gelmini and P.~Gondolo, {\em Phys. Rev.} {\bf
  D74} (2006) 023510; G.~Gelmini and P.~Gondolo, A.~Soldatenko and
  C.~Yaguna, {\em Phys. Rev.} {\bf D74} (2006) 083514.
%
\bibitem{nonuni}H.~Baer, V.~Barger, P.~Huang, D.~Mickelson,
  M.~Padeffke-Kirkland and X.~Tata, {\em Phys. Rev.} {\bf D91} (2015)
  075005.
%
\bibitem{wz} H.~Baer, V.~Barger, S.~Kraml, A.~Lessa, W.~Sreethawong and
X.~Tata, {\em J. High Energy Phys.} {\bf 1203} (2012) 092.
%
\bibitem{wh} H.~Baer, V.~Barger, A.~Lessa, W.~Sreethawong and X.~Tata,
  {\em Phys. Rev.} {\bf D85} (2012) 055022. 
%
\bibitem{lhcwzwh}V.~Khachatryan {\it et al.} (CMS Collaboration)
\epjc{74}{2014}{3036} and {\em Phys. Rev.} {\bf D90} (2014) 092007;
G.~Aad {\it et al.} (ATLAS Collaboration)
\jhep{1405}{2014}{071} and \epjc{75}{2015}{208}.
%
\bibitem{lykken} See {\it e.g.} A.~Strumia, {\em J. High Energy Phys.} {\bf
  1104} (2011) 073; J.~Lykken and M.~Spiropulu,
 {\it  Sci. Am.} {\bf 310N5} (2014) 5,  36. See also, N.~Craig,
  arXiv:1309.0528, for an independent assessment of the state of SUSY in
  light of LHC8 data. 
%
\end{thebibliography}
\end{document}